%% file: opinion.tex
\documentclass[final,11pt,5p,twocolumn,sort&compress]{elsarticle}

\usepackage{amsfonts,amsmath,amssymb}
\usepackage{bbm}
\usepackage{caption,subcaption, epstopdf}

\journal{Annual Reviews in Control}

\def\diag{\mathop{\rm diag}\nolimits}

\def\la{\lambda}
\def\La{\Lambda}

\def\r{\mathbb R}
\def\c{\mathbb C}

\def\g{\mathcal G}

\newcommand{\dfb}{\stackrel{\Delta}{=}}

\def\be{\begin{equation}}
\def\ee{\end{equation}}
\def\ben{\begin{equation*}}
\def\een{\end{equation*}}

\newtheorem{thm}{Theorem}
\newtheorem{lem}[thm]{Lemma}

\newtheorem{cor}[thm]{Corollary}
\newdefinition{defn}{Definition}
\newdefinition{exam}{Example}

\def\ones{\mathbbm{1}}
\def\re{\mathop{\rm Re}\nolimits}
\def\imath{\mathrm{\bf i}}
\def\rk{\mathop{\rm rank}\nolimits}

\begin{document}

\begin{frontmatter}

\title{A Tutorial on Modeling and Analysis of Dynamic Social Networks.~Part I\tnoteref{t}}
\tnotetext[t]{The paper is supported by Russian Science Foundation (RSF) grant 14-29-00142 hosted by IPME RAS.}


\author[TUD,IPME,ITMO]{Anton V. Proskurnikov\corref{mycorrespondingauthor}}
\ead{anton.p.1982@ieee.org}

\author[CNR]{Roberto Tempo}
\cortext[mycorrespondingauthor]{Corresponding author}

\address[TUD]{Delft Center for Systems and Control, Delft University of Technology, Delft, The Netherlands}
\address[IPME]{Institute of Problems of Mechanical Engineering of the Russian Academy of Sciences (IPME RAS), St. Petersburg, Russia}
\address[ITMO]{ITMO University, St. Petersburg, Russia}
\address[CNR]{CNR-IEIIT, Politecnico di Torino, Torino, Italy}

\begin{abstract}
In recent years, we have observed a significant trend  towards filling the gap between social network analysis and control. This trend was enabled by the introduction of new mathematical models
describing dynamics of social groups, the advancement in complex networks theory and multi-agent systems,
and the development of modern computational tools for big data analysis.
The aim of this tutorial is to highlight a novel chapter of control theory, dealing with applications to social systems, to the attention of the broad research community. This paper is the first part of the tutorial, and it is focused on the most classical models of social dynamics and on their relations to the recent achievements in multi-agent systems.
\end{abstract}


\end{frontmatter}


\input 1intro
\input 2prelim

\input 3french

\input 4abelson
\input 5stubborn

\section*{Concluding Remarks and Acknowledgements}

In the first part of the tutorial, we have discussed several dynamic models, in continuous and discrete-time, for opinion formation and evolution. A rigorous analysis of the convergence and stability properties of these models is provided, the relations with recent results on multi-agent systems are discussed.
Due to the page limit, we do not discuss some important ``scalability'' issues, regarding the applicability of the models to large-scale social networks.
Among such issues are the algorithmic analysis of convergence conditions~\cite{JarvisSheer:2000}, the models' convergence rates~\cite{CaoMorse:08Part2,OlshevskyTsitsi:11,GhaderiSrikant:2014,PiraniSundaram:16,ProTempoCao16-2} and experimental validation on big data~\cite{Kerckhove:16}.
An important open problem is to find the relation between the behavior of opinions in the models from Sections~\ref{sec.french}-\ref{sec.fj} and the structure of
\emph{communities} or \emph{modules} in the network's graph~\cite{FortunatoReview:2010}.

In the second part of the tutorial, we are going to discuss advanced more advanced models of opinion formation, based on the ideas of bounded confidence~\cite{Krause:2002,Blondel:2010,MorarescuGirard:2011}, antagonistic interactions~\cite{Altafini:2012,Altafini:2013,ShiBarasJohansson:16} and asynchronous gossip-based interactions~\cite{DeffuantWeisbuch:2000,FrascaIshiiTempo:2015}. Future perspectives of systems and control in social and behavioral sciences will also be discussed.

The authors are indebted to Prof. Noah Friedkin, Prof. Francesco Bullo, Dr. Paolo Frasca and the anonymous reviewers for their invaluable comments.

\bibliographystyle{elsarticle-num}
\bibliography{social,networks,consensus}

\end{document}

%% file: 1intro.tex
\section{Introduction}

The 20th century witnessed a crucial paradigm shift in social and behavioral sciences, which
can be described as ``moving from the description of social bodies to dynamic problems of changing group life''~\citep{Lewin:1947}. Unlike individualistic approaches, focused on
individual choices and interests of social actors, the emerging theories dealt with structural properties
of social groups, organizations and movements, focusing on social relations (or ties) among their members.

A breakthrough in the analysis of social groups was enabled by introducing a quantitative method for describing
social relations, later called~\emph{sociometry}~\citep{Moreno1934,Moreno1951}. The pioneering work~\citep{Moreno1934}
introduced an important graphical tool of \emph{sociogram}, that is, ``a graph that visualizes the underlying
structure of a group and the position each individual has within it''~\citep{Moreno1934}.
The works~\citep{Moreno1934,Moreno1951} also broadly used the term ``network'', meaning a group of individuals that are ``bound together'' by some long-term relationships. Later, the term \emph{social network} was coined, which denotes a structure, constituted by \emph{social actors} (individuals or organizations) and \emph{social ties} among them. Sociometry has given birth to the interdisciplinary science of Social Network Analysis (SNA)~\citep{WassermanFaustBook,Scott_Handbook2000,Freeman2004,HandBookSNA2011},
extensively using mathematical methods and algorithmic tools to study structural properties of social networks and social movements~\citep{DianiAdamBook}. SNA is closely related to economics~\citep{EasleyKleinberg,JacksonBook2008}, political studies~\citep{KnokeBook:1990}, medicine and health care~\citep{OMalleyMarsden:2008}.
The development of SNA has inspired many important concepts of modern network theory ~\citep{Strogatz:2001,NewmanReview:2003,NewmanBarabasiBook} such as e.g. cliques and communities, centrality measures, small-world network, graph's density and clustering coefficient.

On a parallel line of research, Norbert Wiener introduced the general science of Cybernetics~\citep{Wiener1948,WienerBook-Socio} with the objective to unify systems, control and information theory. Wiener believed that this new science should become a powerful tool in studying social processes, arguing that ``society can only be understood through a study of the messages and communication facilities which belong to it''~\citep{WienerBook-Socio}. Confirming Wiener's ideas, the development of social sciences in the 20th century has given birth to a new chapter of sociology, called ``sociocybernetics''~\citep{Geyer:1995} and led to the increasing comprehension that ``the foundational problem of sociology is the coordination and control of social systems''~\citep{Friedkin:2015}. However, the realm of social systems has remained almost untouched by modern control theory
in spite of the tremendous progress in control of complex large-scale systems~\citep{MurrayReport,SamadReport,IFACAgenda2017}.

The gap between the well-developed theory of SNA and control can be explained by the lack of mathematical models, describing social dynamics, and tools for quantitative analysis and numerical simulation of large-scale social groups. While many natural and engineered networks exhibit ``spontaneous order'' effects~\cite{StrogatzSync} (consensus, synchrony and other regular collective behaviors), social communities are often featured by highly ``irregular'' and sophisticated dynamics. Opinions of individuals and actions related to them often fail to reach consensus but rather exhibit persistent disagreement, e.g. clustering or cleavage~\citep{Friedkin:2015}. This requires to develop mathematical models that are sufficiently ``rich'' to capture the behavior of social actors but are also ``simple'' enough to be rigorously analyzed. Although various aspects of ``social'' and ``group'' dynamics have been studied in the sociological literature~\citep{Lewin:1947,SorokinBook1947}, mathematical methods of SNA have focused on graph-theoretic properties of social networks, paying much less attention to dynamics over them. The relevant models have been mostly confined to very special processes, such as e.g. random walks, contagion and percolation~\citep{NewmanReview:2003,NewmanBarabasiBook}.

The recent years have witnessed an important tendency towards filling the gap between SNA and dynamical systems, giving rise to new theories of Dynamical Social Networks Analysis (DSNA)~\citep{BreigerCarley_DSNA2003} and \emph{temporal} or \emph{evolutionary} networks~\citep{HolmeBook:2013,AggarwalReview:2014}. Advancements in statistical physics have given rise to a new science of \emph{sociodynamics}~\citep{Weidlich:2005,Castellano:2009}, which stipulates analogies between social communities and physical systems. Besides theoretical methods for analysis of complex social processes, software tools for big data analysis have been developed, which enable an investigation of Online Social Networks such as Facebook and Twitter and dynamical processes over them~\citep{ArnaboldiBook_OSN}.

Without any doubt, applications of multi-agent and networked control to social groups will become a key component of the emerging science on dynamic social networks. Although the models of social processes have been suggested in abundance~\citep{Mason:2007,Castellano:2009,AcemogluDahleh:2011,XiaWangXuan:2011,Friedkin:2015}, only a few of them
have been rigorously analyzed from the system-theoretic viewpoint. Even less attention has been paid to their experimental validation, which requires to develop rigorous identification methods. A branch of control theory, addressing problems from social and behavioral sciences, is very young, and its contours are still blurred.
 Without aiming to provide a complete and exhaustive survey of this novel area at its dawn, this tutorial focuses on the most ``mature''
dynamic models and on the most influential mathematical results, related to them. These models and results are mainly concerned with \emph{opinion formation} under social influence.

This paper, being the first part of the tutorial, introduces preliminary mathematical concepts and considers the four models of opinion evolution, introduced in 1950-1990s (but rigorously examined only recently): the models by French-DeGroot, Abelson, Friedkin-Johnsen and Taylor. We also discuss the relations between these models and modern multi-agent control, where some of them have been subsequently rediscovered. In the second part of the tutorial more advanced models of opinion evolution, the current trends and novel challenges for systems and control in social sciences will be considered.

The paper is organized as follows. Section~\ref{sec.prelim} introduces some preliminary concepts, regarding multi-agent networks, graphs and matrices. In Section~\ref{sec.french} we introduce the French-DeGroot model and discuss its relation to multi-agent consensus. Section~\ref{sec.abelson} introduces a continuous-time counterpart of the French-DeGroot model, proposed by Abelson; in this section the Abelson diversity problem is also discussed. Sections~\ref{sec.taylor} and~\ref{sec.fj} introduce, respectively, the Taylor and Friedkin-Johnsen models, describing opinion formation in presence of stubborn and prejudiced agents. 

%% file: 2prelim.tex
\section{Opinions, Agents, Graphs and Matrices}\label{sec.prelim}

In this section, we discuss several important concepts, broadly used throughout the paper.

\subsection{Approaches to opinion dynamics modeling}

In this tutorial, we primarily deal with models of \emph{opinion dynamics}. As discussed in~\citep{Friedkin:2015}, individuals' opinions stand for their \emph{cognitive orientations} towards some objects (e.g. particular issues, events or other individuals), for instance, displayed \emph{attitudes}~\cite{Abelson:1964,HunterDanesCohenBook_1984,Kaplowitz:1992} or subjective certainties of belief~\cite{Halpern:1991}.  Mathematically, opinions are just scalar or vector quantities associated with social actors.

Up to now, system-theoretic studies on opinion dynamics have primarily focused on models with \emph{real-valued} (``continuous'') opinions, which can attain continuum of values and are treated as some quantities of interest, e.g. subjective probabilities~\cite{DeGroot,Scaglione:2013}. These models obey systems of ordinary differential or difference equations and can be examined by conventional control-theoretic techniques. A discrete-valued scalar opinion is often associated with some action or decision taken by a social actor, e.g. to support some movement or abstain from it and to vote for or against a bill~\citep{Castellano:2009,Weidlich:1971,Clifford:1973,Granovetter:1978,Holley:1975,Sznajd:2000,Yildiz:2013}. A multidimensional discrete-valued opinion may be treated as a set of cultural traits~\cite{Axelrod:1997}. Analysis of discrete-valued opinion dynamics usually require techniques from advanced probability theory that are mainly beyond the scope of this tutorial.

Models of social dynamics can be divided into two major classes: macroscopic and microscopic models.
Macroscopic models of opinion dynamics are similar in spirit to models of continuum mechanics, based on Euler's formalism; this
approach to opinion modeling is also called Eulerian~\cite{CanutoFagnani:2012,MirtaJiaBullo:2014} or statistical~\cite{Weidlich:1971}.
Macroscopic models describe how the \emph{distribution} of opinions (e.g. the vote preferences on some election or referendum) evolves over time. The statistical approach is typically used in ``sociodynamics''~\cite{Weidlich:2005}  and evolutionary game theory~\cite{SmithBook1982_EvolGames,EasleyKleinberg} (where the ``opinions'' of players stand for their strategies);
some of macroscopic models date back to 1930-40s~\cite{Rashevsky:1939,Rashevsky1947}.

Microscopic, or agent-based, models of opinion formation describes how opinions of individual social actors, henceforth called \emph{agents}, evolve. There is an analogy between the microscopic approach, also called \emph{aggregative}~\citep{Abelson:1967},
and the Lagrangian formalism in mechanics~\cite{CanutoFagnani:2012}. Unlike statistical models, adequate for very large groups (mathematically, the number of agents goes to infinity), agent-based models can describe both small-size and large-scale communities.

 With the aim to provide a basic introduction to social dynamics modeling and analysis, this tutorial is confined to \emph{agent-based} models with \emph{real-valued} scalar and vector opinions, whereas other models are either skipped or mentioned briefly. All the models, considered in this paper, deal with an idealistic closed community, which is neither left
by the agents nor can acquire new members. Hence the size of the group, denoted by $n\ge 2$, remains unchanged.

\subsection{Basic notions from graph theory}

Social interactions among the agents are described by weighted (or valued) directed graphs. We introduce only basic definitions regarding graphs and their properties; a more detailed exposition and examples of specific graphs can be found in textbooks on graph theory, networks or SNA, e.g.~\citep{HararyBook:1965,TutteBook,WassermanFaustBook,EasleyKleinberg,JacksonBook2008,BulloBook-Online}.
The reader familiar with graph theory and matrix theory may skip reading the remainder of this section.

Henceforth the term ``graph'' strands for a directed graph (digraph), formally defined as follows.
\begin{defn}(Graph)
A graph is a pair $\g=(V,E)$, where $V=\{v_1,\ldots,v_n\}$ and $E\subseteq V\times V$ are finite sets.
The elements $v_i$ are called \emph{vertices} or \emph{nodes} of $\g$ and the elements of $E$ are referred to as its \emph{edges} or \emph{arcs}.
\end{defn}

Connections among the nodes are conveniently encoded by the graph's \emph{adjacency} matrix $A=(a_{ij})$. In graph theory, the arc $(i,j)$ usually corresponds to the positive entry
$a_{ij}>0$. In multi-agent control~\citep{RenBeardBook,RenCaoBook} and opinion formation modeling it is however convenient\footnote{This definition is motivated by consensus protocols and other models of opinion dynamics, discussed in Sections~\ref{sec.french}-\ref{sec.fj}. It allows to identify the entries of an adjacency matrix with the
influence gains, employed by the opinion formation model.} to identify the arc $(i,j)$ with the entry $a_{ji}>0$.

\begin{defn}(Adjacency matrix)
Given a graph $\g=(V,E)$, a \emph{nonnegative} matrix $A=(a_{ij})_{i,j\in V}$ is \emph{adapted} to $\g$ or is a \emph{weighted adjacency matrix} for $\g$ if $(i,j)\in E$
when $a_{ji}>0$ and $(i,j)\not\in E$ otherwise.
\end{defn}

\begin{defn}(Weighted graph)
A weighted graph is a triple $\g=(V,E,A)$, where $(V,E)$ is a graph and $A$ is a weighted adjacency matrix for it.
\end{defn}

Any graph $(V,E)$ can be considered as a weighted graph by assigning to it a binary adjacency matrix
\[
A=(a_{ij})_{i,j\in V},\quad a_{ij}=\begin{cases}
1,\quad (j,i)\in E\\
0,\quad\text{otherwise}.
\end{cases}
\]
On the other hand, any nonnegative matrix $A=(a_{ij})_{i,j\in V}$ is adapted to the unique graph $\g[A]=(V,E[A],A)$. Typically, the nodes are in one-to-one correspondence with the agents and $V=\{1,\ldots,n\}$.

\begin{defn}(Subgraph)
The graph $\g=(V,E)$ \emph{contains} the graph $\g'=(V',E')$, or $\g'$ is a \emph{subgraph} of $\g$, if $\emptyset\ne V'\subseteq V$ and $E'\subseteq(V'\times V')\cap E$.
\end{defn}

Simply speaking, the subgraph is obtained from the graph by removing some arcs and some nodes.

\begin{defn}(Walk)
A \emph{walk} of length $k$ connecting node $i$ to node $i'$ is a sequence of nodes $i_0,\ldots,i_k\in V$, where $i_0=i$, $i_k=i'$ and adjacent nodes are connected by arcs: $(i_{s-1},i_{s})\in E$ for any $s=1,\ldots,k$. A walk from a node to itself is a \emph{cycle}. A trivial cycle of length $1$ is called a \emph{self-loop} $(v,v)\in E$. A walk without self-intersections ($i_p\ne i_q$ for $p\ne q$) is a \emph{path}.
\end{defn}

It can be shown that in a graph with $n$ nodes the shortest walk between two different nodes (if such a walk exists) has the length $\leq n-1$ and the shortest cycle from a node to itself has the length $\leq n$.

\begin{defn}(Connectivity)
A node connected by walks to all other nodes in a graph is referred to as a \emph{root} node. A graph is called \emph{strongly connected} or \emph{strong} if a walk between any two nodes exists (and hence each node is a root). A graph is \emph{quasi-strongly connected} or \emph{rooted} if at least one root exists.
\end{defn}

\begin{figure}
\begin{subfigure}[b]{0.49\columnwidth}
\center
\includegraphics[width=0.5\columnwidth]{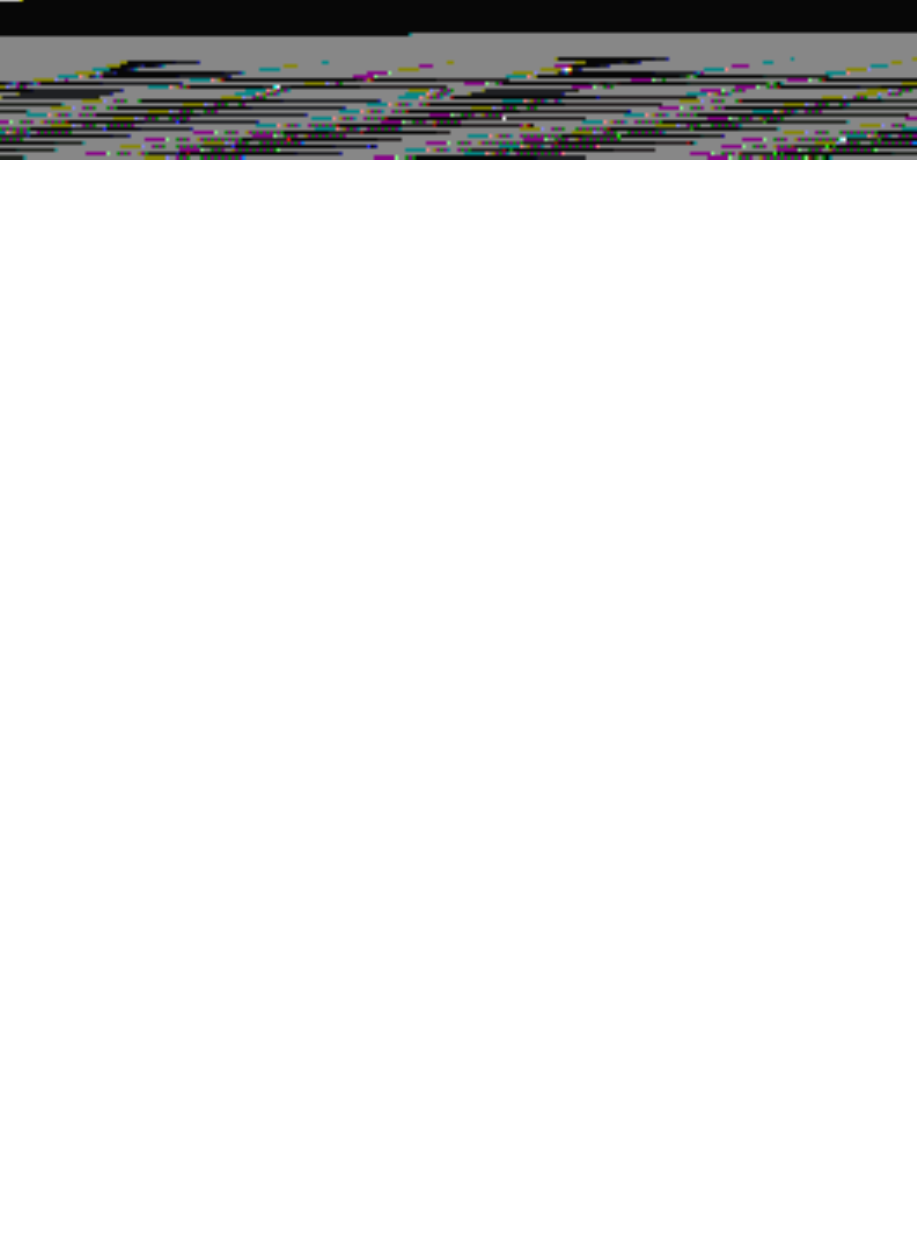}
\caption{}
\end{subfigure}
\begin{subfigure}[b]{0.49\columnwidth}
\center
\includegraphics[width=0.5\columnwidth]{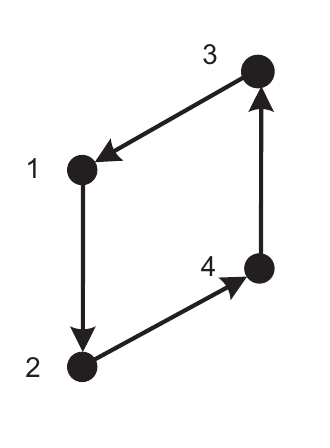}
\caption{}
\end{subfigure}
\caption{Examples of graphs: (a) a directed tree with root $4$; (b) a cyclic graph of period $4$}\label{fig.tree}
\end{figure}
\begin{figure}
\begin{subfigure}[b]{0.49\columnwidth}
\center
\includegraphics[width=0.75\columnwidth]{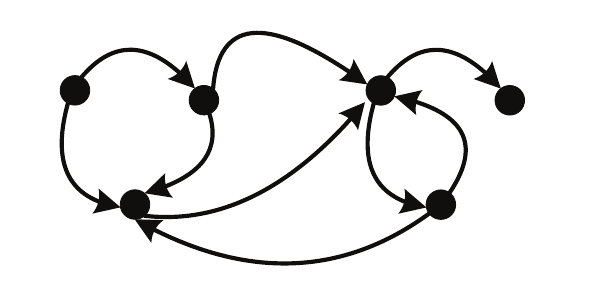}
\caption{}
\end{subfigure}\hfill
\begin{subfigure}[b]{0.49\columnwidth}
\center
\includegraphics[width=0.75\columnwidth]{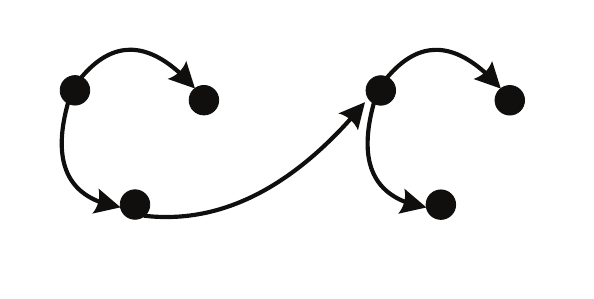}
\caption{}
\end{subfigure}
\caption{A quasi-strongly connected graph (a) and one of its directed spanning trees (b)}\label{fig.tree1}
\end{figure}

The ``minimal'' quasi-strongly connected graph is a \emph{directed tree} (Fig.~\ref{fig.tree}a), that is, a graph with only one root node, from where any other node is reachable via only one walk. A directed \emph{spanning tree} in a graph $\g$ is a directed tree, contained by the graph $\g$ and including all of its nodes (Fig.~\ref{fig.tree1}b).
It can be shown~\citep{RenBeardBook} that a graph has at least one directed spanning tree if and only if it is quasi-strongly connected.
Nodes of a graph without directed spanning tree are covered by several directed trees, or \emph{spanning forest}~\citep{ChebotarevAgaev:2002}.

\begin{defn}(Components)\label{def.component}
A strong subgraph $\g'$ of the graph $\g$ is called a strongly connected (or strong) \emph{component}, if it is not contained by any larger strong subgraph. A strong component that has no incoming arcs from other components is called \emph{closed}.
\end{defn}

Any node of a graph is contained in one and only one strong component. This component may correspond with the whole graph; this holds if and only if the graph is strong.
If the graph is not strongly connected, then it contains two or more strong components, and at least one of them is \emph{closed}.
A graph is quasi-strongly connected if and only if this closed strong component is unique; in this case, any node of this strong component is a root node.

\begin{figure}
\begin{subfigure}[b]{0.49\columnwidth}
\center
\includegraphics[width=0.9\columnwidth]{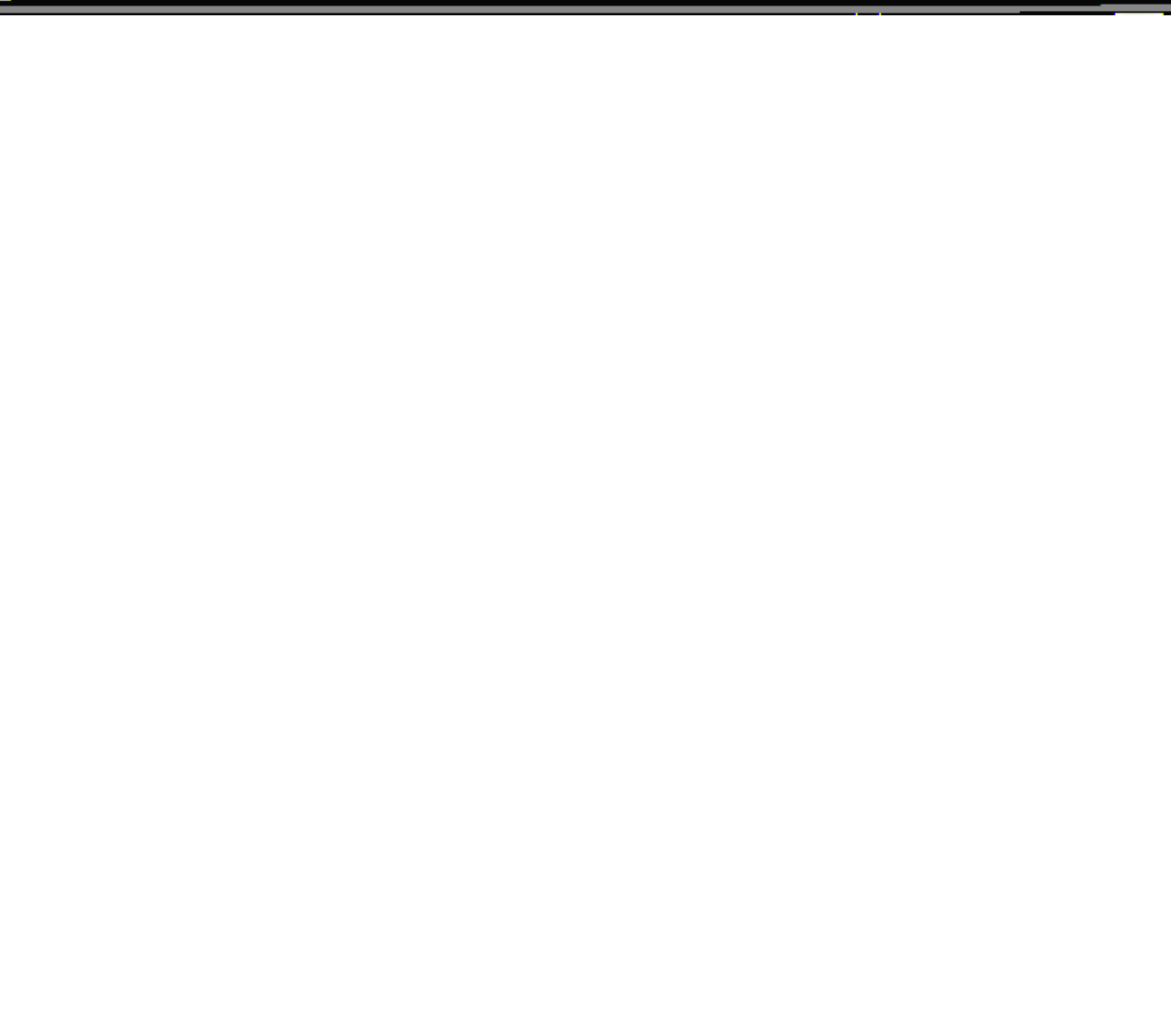}
\caption{}
\end{subfigure}\hfill
\begin{subfigure}[b]{0.49\columnwidth}
\center
\includegraphics[width=0.9\columnwidth]{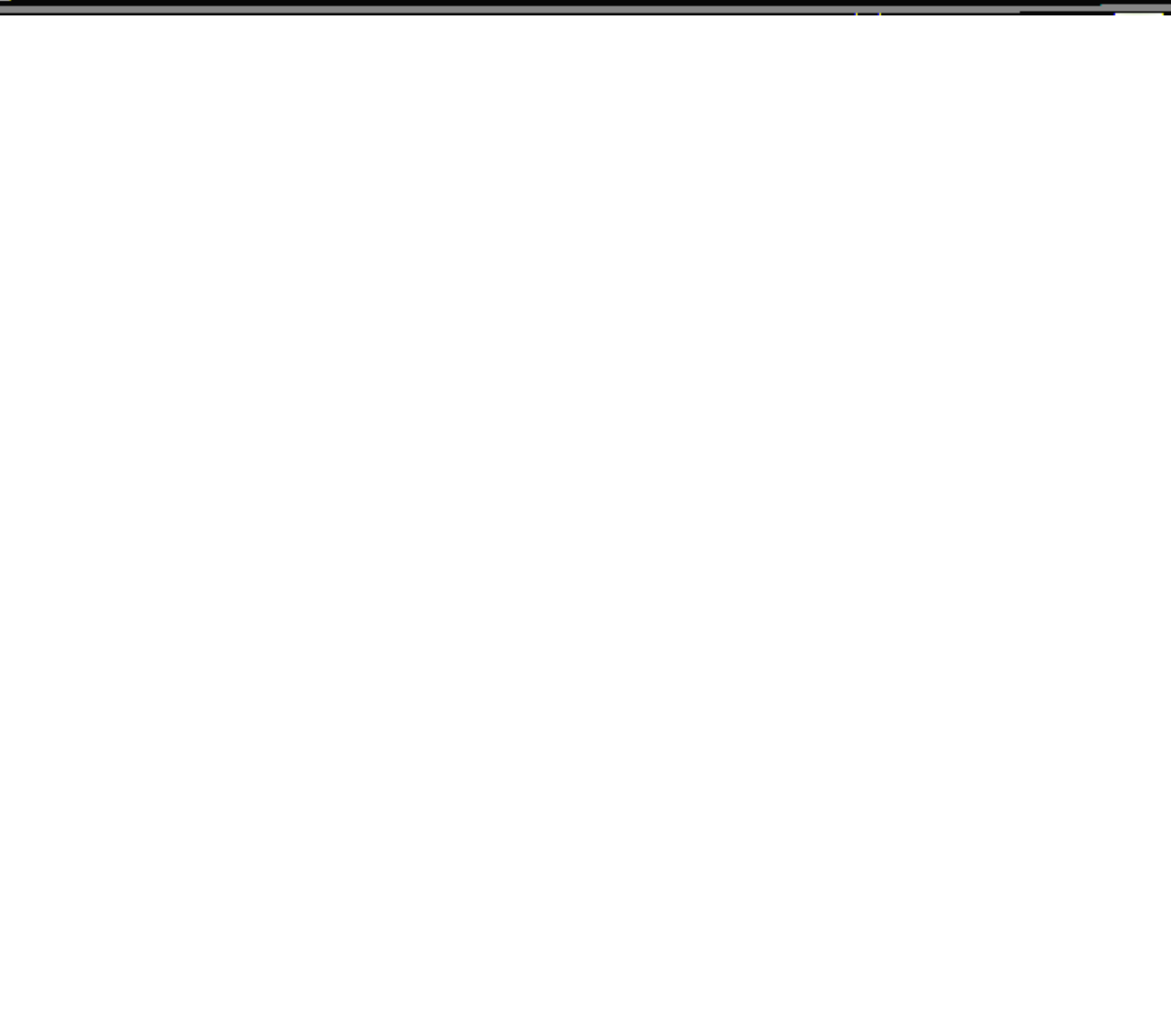}
\caption{}
\end{subfigure}
\caption{Strong components of a rooted graph (a) and a graph without roots (b)}\label{fig.strong}
\end{figure}

Definition~\ref{def.component} is illustrated by Fig.~\ref{fig.strong}, showing two graphs with the same structure of strong components.
The graph in Fig.~\ref{fig.strong}a has the single root node $4$, constituting its own strong component, all other strong components are not closed. The graph in Fig.~\ref{fig.strong}b has two closed strong components $\{4\}$ and $\{5,6,\ldots,10\}$.

\subsection{Nonnegative matrices and their graphs}

In this subsection we discuss some results from matrix theory, regarding nonnegative matrices~\citep{BermanPlemmons1979,HornJohnsonBook1985,Meyer2000Book,GantmacherVol2}.

\begin{defn}(Irreducibility)
A nonnegative matrix $A$ is \emph{irreducible} if $\g[A]$ is strongly connected.
\end{defn}
\begin{thm}(Perron-Frobenius)\label{thm.perron}
The spectral radius $\rho(A)\ge 0$ of a nonnegative matrix $A$ is an eigenvalue of $A$,
for which a real nonnegative eigenvector exists
\[
Av=\rho(A)v\;\;\text{for some}\;\; v=(v_1,\ldots,v_n)^{\top}\ne 0,\,v_i\ge 0.
\]
If $A$ is irreducible, then $\rho(A)$ is a simple eigenvalue and $v$ is strictly positive $v_i>0\,\forall i$.
\end{thm}
Obviously, Theorem~\ref{thm.perron} is also applicable to the transposed matrix $A^{\top}$, and thus $A$ also has a left nonnegative eigenvector $w^{\top}$, such that $w^{\top}A=\rho(A)w^{\top}$.

Besides $\rho(A)$, a nonnegative matrix can have other eigenvalues $\la$ of maximal modulus $|\la|=\rho(A)$. These eigenvalues
have the following property~\citep[Ch.XIII]{GantmacherVol2}.
\begin{lem}\label{lem.mult}
If $A$ is a nonnegative matrix and $\la\in\mathbb{C}$ is its eigenvalue with $|\la|=\rho(A)$, then the algebraic and geometric multiplicities of $\la$ coincide
(that is, all Jordan blocks corresponding to $\la$ are trivial).
\end{lem}

For an \emph{irreducible} matrix, the eigenvalues of maximal modulus are always simple and have the form $\rho(A)e^{2\pi m\imath/h}$, where $h\ge 1$ is some integer and $m=0,1,\ldots,h-1$.
This fundamental property is proved e.g. in~\citep[Sections 8.4 and 8.5]{HornJohnsonBook1985} and~\citep[Section 8.3]{Meyer2000Book}.
\begin{thm}\label{thm.spectrum}
Let an \emph{irreducible} matrix $A$ have $h\ge 1$ different eigenvalues $\la_1,\ldots,\la_h$ on the circle $\{\la\in\mathbb{C}:|\la|=\rho(A)\}$.
Then, the following statements hold:
\begin{enumerate}
\item each $\la_i$ has the algebraic multiplicity $1$;
\item $\{\la_1,\ldots,\la_h\}$ are roots of the equation $\la^h=r^h$;
\item if $h=1$ then all entries of the matrix $A^k$ are strictly positive when $k$ is sufficiently large;
\item if $h>1$, the matrix $A^k$ may have positive diagonal entries only when $k$ is a multiple of $h$.
\end{enumerate}
\end{thm}
\begin{defn}(Primitivity)
An irreducible nonnegative matrix $A$ is \emph{primitive} if $h=1$, i.e. $\la=\rho(A)$ is the only eigenvalue of maximal modulus;
otherwise, $A$ is called~\emph{imprimitive} or~\emph{cyclic}.
\end{defn}

It can be shown via induction on $k=1,2,\ldots$ that if $A$ is a nonnegative matrix and $B=(b_{ij})=A^k$, then $b_{ij}>0$ if and only if
in $\g[A]$ there exists a walk of length $k$ from $j$ to $i$. In particular, the diagonal entry $(A^k)_{ii}$ is positive if and only if a \emph{cycle} of length $k$ from node $i$ to itself exists. Hence, cyclic irreducible matrices correspond to \emph{periodic} strong graphs.
\begin{defn}(Periodicity)
A graph is \emph{periodic} if it has at least one cycle and the length of any cycle is divided by some integer $h>1$.
The maximal $h$ with such a property is said to be the \emph{period} of the graph.
Otherwise, a graph is called \emph{aperiodic}.
\end{defn}

The simplest example of a periodic graph is a \emph{cyclic} graph (Fig.~\ref{fig.tree}b).
Any graph with self-loops is aperiodic. Theorem~\ref{thm.spectrum} implies the following corollary.
\begin{cor}\label{cor.periodic}
An irreducible matrix $A$ is primitive if and only if $\g[A]$ is aperiodic. Otherwise, $\g[A]$ is periodic with period $h$, where
$h>1$ is the number of eigenvalues of the maximal modulus $\rho(A)$.
\end{cor}

Many models of opinion dynamics employ \emph{stochastic} and \emph{substochastic} nonnegative matrices.

\begin{defn}(Stochasticity and substochasticity)
A nonnegative matrix $A$ (not necessarily square) is called \emph{stochastic} if all its rows sum to $1$ (i.e. $\sum_j a_{ij}=1\,\forall i$) and \emph{substochastic} if
the sum of each row is no greater than $1$ (i.e. $\sum_j a_{ij}\le 1\,\forall i$).
\end{defn}

The Gershgorin Disc Theorem~\cite[Ch.~6]{HornJohnsonBook1985} implies that $\rho(A)\le 1$ for any square substochastic matrix $A$. If $A$ is stochastic then $\rho(A)=1$ since $A$ has an eigenvector
of ones $\ones\dfb (1,\ldots,1)^{\top}$: $A\ones=\ones$. A substochastic matrix $A$, as shown in~\cite{Parsegov2017TAC}, either is Schur stable or has a \emph{stochastic} submatrix $A'=(a_{ij})_{i,j\in V'}$, where $V'\subseteq\{1,\ldots,n\}$; an \emph{irreducible} substochastic matrix is either stochastic or Schur stable~\cite{Meyer2000Book,Parsegov2017TAC}.

\subsection{M-matrices and Laplacians of weighted graphs}

In this subsection we introduce the class of \emph{M-matrices}\footnote{The term ``M-matrix'' was suggested by A. Ostrowski in honor of Minkowski, who studied such matrices in early 1900s.}~\cite{BermanPlemmons1979,Meyer2000Book} that are closely related to nonnegative matrices and have some important properties.
\begin{defn}(M-matrix)
A square matrix $Z$ is an \emph{M-matrix} if it admits a decomposition $Z=sI-A$, where $s\ge\rho(A)$ and the matrix $A$ is nonnegative.
\end{defn}

For instance, if $A$ is a substochastic matrix then $Z=I-A$ is an M-matrix.
Another important class of M-matrices is given by the following lemma.
\begin{lem}\label{lem.m-matrix}
Let $Z=(z_{ij})$ satisfies the following two conditions: 1) $z_{ij}\le 0$ when $i\ne j$; 2) $z_{ii}\ge\sum_{j\ne i}|z_{ij}|$. Then, $Z$ is an M-matrix; precisely, $A=sI-Z$ is nonnegative and $\rho(A)\le s$ whenever $s\ge\max_iz_{ii}$.
\end{lem}
Indeed, if $s\ge\max_iz_{ii}$ then $A=sI-Z$ is nonnegative and $\rho(A)\le\max_{i}(s-z_{ii}+\sum_{j\ne i}|z_{ij}|)\le s$ thanks to the Gershgorin Disc Theorem~\cite{HornJohnsonBook1985}.

Noticing that the eigenvalues of $Z$ and $A$ are in one-to-one correspondence $\la\mapsto s-\la$ and using
Theorem~\ref{thm.perron} and Lemma~\ref{lem.mult}, one arrives at the following result.
\begin{cor}\label{cor.m-matrix}
Any M-matrix $Z=sI-A$ has a real eigenvalue $\la_0=s-\rho(A)\ge 0$, whose algebraic and geometric multiplicities coincide.
For this eigenvalue there exist nonnegative right and left eigenvectors $v$ and $p$: $Zv=\la_0v$, $p^{\top}Z=\la_0p^{\top}$.
These vectors are positive if the graph $\g[-Z]$ is strongly connected.
For any other eigenvalue $\la$ one has $\re\la>\la_0$, and hence $Z$ is non-singular ($\det Z\ne 0$) if and only if $s>\rho(A)$.
\end{cor}

Non-singular $M$-matrices are featured by the following important property~\cite{BermanPlemmons1979,Meyer2000Book}.
\begin{lem}\label{lem.m-matrix-inverse}
Let $Z=sI-A$ be a non-singular $M$-matrix, i.e. $s>\rho(A)$. Then $Z^{-1}$ is \emph{nonnegative}.
\end{lem}

An example of a \emph{singular} M-matrix is the \emph{Laplacian} (or \emph{Kirchhoff}) matrix of a weighted graph~\citep{Godsil,TutteBook,AgaevChe:2005}.
\begin{defn}(Laplacian)
Given a weighted graph $\g=(V,E,A)$, its Laplacian matrix is defined by
\be\label{eq.laplacian}
L[A]=(l_{ij})_{i,j\in V},\;\text{where}\;\; l_{ij}=
\begin{cases}
-a_{ij},&i\ne j\\
\sum\limits_{j\ne i}a_{ij}, &i=j.
\end{cases}
\ee
\end{defn}

The Laplacian is an M-matrix due to Lemma~\ref{lem.m-matrix}. Obviously, $L[A]$ has the eigenvalue $\la_0=0$ since $L[A]\ones_n=0$, where $n$ is the dimension of $A$. The zero eigenvalue is simple if and only if the graph $\g[A]$ has a directed spanning tree (quasi-strongly connected).
\begin{lem}\label{lem.laplacian}
For an arbitrary nonnegative square matrix $A$ the following conditions are equivalent
\begin{enumerate}
\item $0$ is an algebraically simple eigenvalue of $L[A]$;
\item if $L[A]v=0,\,v\in\r^n$ then $v=c\ones_n$ for some $c\in\r$ (e.g. $0$ is a geometrically simple eigenvalue);
\item the graph $\g[A]$ is quasi-strongly connected.
\end{enumerate}
\end{lem}
The equivalence of statements 1 and 2 follows from Corollary~\ref{cor.m-matrix}. The equivalence of statements 2 and 3 was in fact proved in~\citep{Abelson:1964} and rediscovered in recent papers~\citep{Ren:05,LinFrancis:05}. A more general relation between the kernel's dimension $\dim\ker L[A]=n-\rk L[A]$
and the graph's structure has been established\footnote{As discussed in~\cite[Section 6.6]{BulloBook-Online}, the first studies on the Laplacian's rank date back to 1970s and were motivated by the dynamics of compartmental systems in mathematical biology.} in~\citep{ChebotarevAgaev:2002,AgaevChe:2005}.

%% file: 3french.tex
\section{The French-DeGroot Opinion Pooling}\label{sec.french}

One of the first \emph{agent-based} models\footnote{As was mentioned in Section~\ref{sec.prelim}, a few statistical models of social
systems had appeared earlier, see in particular~\cite{Rashevsky:1939,Rashevsky1947}.} of opinion formation was proposed by the social psychologist French in his influential paper~\citep{French:1956}, binding together SNA and systems theory. Along with its generalization, suggested by DeGroot~\citep{DeGroot} and called ``iterative opinion pooling'', this model describes a simple
procedure, enabling several rational agents to reach consensus~\citep{Lehrer:1976,Wagner:1978,LehrerWagnerBook}; it may also be considered as an algorithm of non-Bayesian \emph{learning}~\citep{GolubJackson:2010,Jadbabaie:2012}. The original goal of French, however, was not to study consensus and learning mechanisms but rather to find a mathematical model for \emph{social power}~\cite{French:1956,FrenchRaven:1959,Friedkin:1986}. An individual's social power in the group is his/her ability to control the group's behavior, indicating thus the \emph{centrality} of the individual's node in the social network. French's work has thus revealed a profound relation between opinion formation and centrality measures.

\subsection{The French-DeGroot model of opinion formation}

The French-DeGroot model describes a discrete-time process of opinion formation in a group of $n$ agents, whose opinions henceforth
 are denoted by $x_1,\ldots,x_n$. First we consider the case of scalar opinions $x_i\in\r$. The key parameter of the model is a stochastic $n\times n$ matrix of \emph{influence weights}
$W=(w_{ij})$. The influence weights $w_{ij}\ge 0$, where $j=1,\ldots,n$ may be considered as some finite resource, distributed by agent $i$ to self and the other agents. Given a positive influence weight $w_{ij}>0$, agent $j$ is able to influence the opinion of agent $i$ at each step of the opinion iteration; the greater weight is assigned to agent $j$, the stronger is its influence on agent $i$.
Mathematically, the vector of opinions $x(k)=(x_1(k),\ldots,x_n(k))^{\top}$ obeys the equation
\be\label{eq.french}
x(k+1)=Wx(k),\quad k=0,1,\ldots.
\ee
which is equivalent to the system of equations
\be\label{eq.french-coordinate}
x_i(k+1)=\sum_{j=1}^nw_{ij}x_j(k),\,\forall i\quad k=0,1,\ldots.
\ee
Hence $w_{ij}$ is the contribution of agent $j$'s opinion at each step of the opinion iteration to the opinion of agent $i$ at its next step. The self-influence weight $w_{ii}\ge 0$ indicates the agent's openness to the assimilation of the others' opinions: the agent with $w_{ii}=0$ is open-minded and completely relies on the others' opinions, whereas the agent with $w_{ii}=1$ (and $w_{ij}=0\,\forall j\ne i$) is a \emph{stubborn} or \emph{zealot} agent, ``anchored'' to its initial opinion~$x_i(k)\equiv x_i(0)$.

More generally, agent's opinions may be \emph{vectors} of dimension $m$, conveniently represented by \emph{rows}  $x^i=(x_{i1},\ldots,x_{im})$. Stacking these rows on top of one another, one obtains an opinion matrix
$X=(x_{il})\in\r^{n\times m}$. The equation~\eqref{eq.french} should be replaced by
\be\label{eq.degroot}
X(k+1)=WX(k),\quad k=0,1,\ldots.
\ee
Every column $x_i(k)=(x_{1i}(k),\ldots,x_{ni}(k))^{\top}\in\r^n$ of $X(k)$, obviously, evolves in accordance with~\eqref{eq.french}.
Henceforth the model~\eqref{eq.degroot} with a general stochastic matrix $W$ is referred to as the \emph{French-DeGroot} model.

\subsection{History of the French-DeGroot model}

\begin{figure}[h]
\center
\includegraphics[width=0.6\columnwidth]{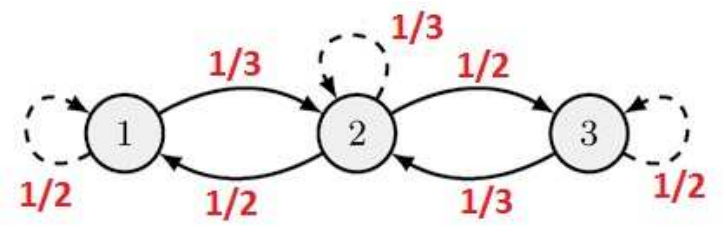}
\caption{An example of the French model with $n=3$ agents}\label{fig.french}
\end{figure}

A special case of the model~\eqref{eq.french} has been introduced by French in his seminal paper~\citep{French:1956}.
This paper first introduces a graph $\g$, whose nodes correspond to the agents; it is assumed that each node has a self-loop. An arc $(j,i)$ exists if agent $j$'s opinion is displayed to agent $i$,
or $j$ ``has \emph{power} over'' $i$. At each stage of the opinion iteration, an agent updates its opinion to the \emph{mean value} of the opinions, displayed to it, e.g. the weighted graph in Fig.~\ref{fig.french} corresponds to the dynamics
\be\label{eq.ex-1}
\begin{bmatrix}
x_1(k+1)\\
x_2(k+1)\\
x_3(k+1)
\end{bmatrix}=
\begin{pmatrix}
1/2 & 1/2 & 0\\
1/3 & 1/3 & 1/3\\
0 & 1/2 & 1/2
\end{pmatrix}
\begin{bmatrix}
x_1(k)\\
x_2(k)\\
x_3(k)
\end{bmatrix}.
\ee
Obviously, the French's model is a special case of equation~\eqref{eq.french}, where the matrix $W$ is adapted to the graph $\g$ and has positive diagonal entries;
furthermore, in each row of $W$ all non-zero entries are equal. Hence each agent \emph{uniformly} distributes influence between itself and
the other nodes connected to it.

French formulated without proofs several conditions for reaching a \emph{consensus}, i.e. the convergence $x_i(k)\xrightarrow[k\to\infty]{} x_*$ of all opinions to a common ``unanimous opinion''~\citep{French:1956} that were later corrected and rigorously proved by Harary~\cite{Harary:1959,HararyBook:1965}. His primary interest was, however, to find a quantitative characteristics of the agent's \emph{social power}, that is, its ability to influence the group's collective opinion $x_*$ (the formal definition will be given in Subsect.~\ref{subsec.power}).

A general model~\eqref{eq.degroot}, proposed by DeGroot~\citep{DeGroot}, takes its origin in applied statistics and has been suggested
as a heuristic procedure to find ``consensus of subjective probabilities''~\citep{EisenbergGale:1959}. Each of $n$ agents (``experts'')
has a vector opinion, standing for an individual (``subjective'') \emph{probability distribution} of $m$ outcomes in some random experiment; the experts' goal is to ``form a distribution which represents, in some sense, a consensus of individual distributions''~\citep{EisenbergGale:1959}. This distribution was defined in~\citep{EisenbergGale:1959} as the unique Nash equilibrium~\citep{Nash:1951} in a special non-cooperative ``Pari-Mutuel'' game (betting on horse races), which can be found by solving a special optimization problem, referred now to as the \emph{Eisenberg-Gale convex program}~\citep{JainVazirani:2010}.
To obtain a simpler algorithm of reaching consensus, a heuristical algorithm was suggested in~\cite{Winkler:1968}, replacing
the convex optimization by a very simple procedure of weighted averaging,
or \emph{opinion pooling}~\citep{Stone:1961}. Developing this approach, the procedure of \emph{iterative} opinion pooling~\eqref{eq.degroot} was suggested in~\citep{DeGroot}. Unlike~\citep{EisenbergGale:1959,Winkler:1968}, the DeGroot procedure was a \emph{decentralized} algorithm: each agent modifies its opinion independently based on the opinions of several ``trusted'' individuals, and there may be no agent aware of the opinions of the whole group. Unlike the French model~\cite{French:1956},
the matrix $W$ can be an arbitrary stochastic matrix and the opinions are vector-valued.

\subsection{Algebraic convergence criteria}

In this subsection, we discuss convergence properties of the French-DeGroot model~\eqref{eq.french}; the properties for the multidimensional model~\eqref{eq.degroot} are the same.

A straightforward computation shows that the dynamics~\eqref{eq.french} is ``non-expansive'' in the sense that
\[
\begin{gathered}
\min_i x_i(0)\le\cdots\le \min_ix_i(k)\le \min_ix_i(k+1),\\
\max_i x_i(0)\ge\cdots\ge \max_ix_i(k)\ge \max_ix_i(k+1)\\
\end{gathered}
\]
for any $k=0,1,\ldots$. In particular, the system~\eqref{eq.french} is always \emph{Lyapunov stable}\footnote{This also follows from Lemma~\ref{lem.mult} since $\rho(W)=1$.}, but this stability is not asymptotic since $W$ always has eigenvalue at $1$.

The first question, regarding the model~\eqref{eq.degroot}, is whether the opinions converge or oscillate.
A more specific problem is convergence to a \emph{consensus}~\citep{DeGroot}.
\begin{defn}(Convergence)
The model~\eqref{eq.french} is \emph{convergent} if for any initial condition $x(0)$ the limit exists
\be\label{eq.x_*}
x(\infty)=\lim_{k\to\infty} x(k)=\lim_{k\to\infty} W^kx(0).
\ee
A convergent model reaches a \emph{consensus} if $x_1(\infty)=\ldots=x_n(\infty)$ for any initial opinion vector $x(0)$.
\end{defn}

The convergence and consensus in the model~\eqref{eq.french} are equivalent, respectively, to
\emph{regularity} and \emph{full regularity}\footnote{Our terminology follows~\citep{GantmacherVol2}. The term ``regular matrix'' sometimes denotes a fully regular matrix~\citep{Seneta} or a primitive matrix~\citep{KemenySnellMarkovBook}.
Fully regular matrices are also referred to as \emph{SIA} (stochastic indecomposable aperiodic) matrices~\citep{Wolfowitz:1963,Jad:03,RenBeardBook,RenCaoBook}.} of the stochastic matrix $W$.
\begin{defn}(Regularity)\label{def.regul}
We call the matrix $W$ \emph{regular} if the limit $W^{\infty}=\lim\limits_{k\to\infty}W^k$ exists and \emph{fully regular} if, additionally, the rows of $W^{\infty}$ are identical (that is, $W^{\infty}=\ones_np^{\top}_{\infty}$ for some $p_{\infty}\in\r^n$).
\end{defn}
Lemma~\ref{lem.mult} entails the following convergence criterion.
\begin{lem}~\citep[Ch.XIII]{GantmacherVol2}\label{lem.gantm}
The model~\eqref{eq.french} is convergent (i.e. $W$ is regular) if and only if $\la=1$ is the only eigenvalue of $W$ on the unit circle $\{\la\in\c:|\la|=1\}$. The model~\eqref{eq.french} reaches consensus (i.e. $W$ is fully regular) if and only if this eigenvalue is simple, i.e. the corresponding eigenspace is spanned by the vector $\ones$.
\end{lem}
Using Theorem~\ref{thm.spectrum}, Lemma~\ref{lem.gantm} implies the equivalence of convergence and consensus when $W$ is irreducible.
\begin{lem}\label{lem.primitivity1}
For an irreducible stochastic matrix $W$ the model~\eqref{eq.french} is convergent if and only if $W$ is primitive, i.e. $W^k$ is a positive matrix for large $k$. In this case consensus is also reached.
\end{lem}

Since an imprimitive irreducible matrix $W$ has eigenvalues $\{e^{2\pi k\imath/h}\}_{k=0}^{h-1}$, where $h>1$,
for almost all\footnote{``Almost all'' means ``all except for a set of zero measure''.} initial conditions the solution of~\eqref{eq.french} oscillates.

\subsection{Graph-theoretic conditions for convergence}

For large-scale social networks, the criterion from Lemma~\ref{lem.gantm} cannot be easily tested. In fact, convergence
of the French-DeGroot model~\eqref{eq.french} does not depend on the weights $w_{ij}$ , but only on the graph $\g[W]$.
In this subsection, we discuss graph-theoretic conditions for convergence and consensus.
Using Corollary~\ref{cor.periodic}, Lemma~\ref{lem.primitivity1} may be reformulated as follows.
\begin{lem}\label{lem.primitivity2}
If the graph $\g=\g[W]$ is strong, then the model~\eqref{eq.french} reaches a consensus if and only if $\g$ is aperiodic.
Otherwise, the model is not convergent and opinions oscillate for almost all $x(0)$.
\end{lem}

Considering the general situation, where $\g[W]$ has more than one strong component, one may easily notice that the evolution of the opinions in any \emph{closed} strong component is independent from the remaining network.
Two different closed components obviously cannot reach consensus for a general initial condition.

This implies that for convergence of the opinions it is necessary that all closed strong components are aperiodic. For reaching a consensus the graph $\g[W]$ should have \emph{the only} closed strong component (i.e. be quasi-strongly connected), which is aperiodic. Both conditions, in fact, appear to be \emph{sufficient}.
\begin{thm}~\citep{DeMarzo:2003,JacksonBook2008}\label{thm.degroot}
The model~\eqref{eq.french} is convergent if and only if all closed strong components in $\g[W]$ are aperiodic. The model~\eqref{eq.french} reaches a consensus if and only if $\g[W]$ is quasi-strongly connected and the only closed strong component is aperiodic.
\end{thm}

As shown in the next subsection, Theorem~\ref{thm.degroot} can be derived from the standard results on the Markov chains convergence~\citep{Seneta}, using the duality between Markov chains and the French-DeGroot opinion dynamics.
Theorem~\ref{thm.degroot} has an important corollary, used in the literature on multi-agent consensus.
\begin{cor}\label{cor.harary}
Let the agents' self-weights be positive $w_{ii}>0\,\forall i$. Then, the model~\eqref{eq.french} is convergent. It reaches a consensus if and only if $\g[W]$ is
quasi-strongly connected (i.e. has a directed spanning tree).
\end{cor}
It should be noted that the existence of a directed spanning tree is in general \emph{not} sufficient for consensus in the case where $W$ has zero diagonal entries.
The second part of Corollary~\ref{cor.harary} was proved\footnote{Formally,~\citep{Harary:1959,HararyBook:1965} address only the French model, however, the proof uses only the diagonal entries' positivity $w_{ii}>0\,\forall i$.} in~\citep{Harary:1959} and included, without proof, in~\citep[Chapter 4]{HararyBook:1965}. Numerous extensions of this result to time-varying matrices $W(k)$~\citep{Tsitsiklis:86,DeMarzo:2003,Blondel:05,RenBeardBook,CaoMorse:08Part1} and more general nonlinear consensus algorithms~\citep{Moro:05,Bliman:06} have recently been obtained.
Some time-varying extensions of the French-DeGroot model, namely, bounded confidence opinion dynamics~\cite{Krause:2002} and dynamics of reflected appraisal~\cite{FriedkinJiaBullo:2016}
will be discussed in Part~II of this tutorial.

\subsection{The dual Markov chain and social power}\label{subsec.power}

Notice that the matrix $W$ may be considered as a matrix of transition probabilities of some Markov chain with $n$ states. Denoting by $p_i(t)$ the probability of being at state $i$ at time $t$, the row  vector $p^{\top}(t)=(p_1(t),\ldots,p_n(t))$ obeys the equation
\be\label{eq.french1-dual}
p(k+1)^{\top}=p(k)^{\top}W,\quad t=0,1,\ldots
\ee
The convergence of~\eqref{eq.french}, that is, regularity of $W$ implies that the probability distribution converges to the limit $p(\infty)^{\top}=\lim_{k\to\infty}p(k)^{\top}=p(0)^{\top}W^{\infty}$. Consensus in~\eqref{eq.french} implies that $p(\infty)=p_{\infty}$, where $p_{\infty}$ is the vector from Definition~\ref{def.regul}, i.e. the Markov chain ``forgets'' its history and convergence to the unique stationary distribution. Such a chain is called \emph{regular} or \emph{ergodic}~\citep{SenetaBook,GantmacherVol2}. The \emph{closed} strong components in $\g[W]$ correspond to \emph{essential classes} of states, whereas the remaining nodes correspond to \emph{inessential} (or non-recurrent) states~\citep{SenetaBook}. The standard ergodicity condition is that the essential class is unique and aperiodic, which is in fact equivalent to the second part of Theorem~\ref{thm.degroot}. The first part of Theorem~\ref{thm.degroot}
states another known fact~\citep{SenetaBook}: the Markov chain always converges to a stationary distribution if and only if all essential classes are aperiodic.

Assuming that $W$ is fully regular, one notices that
\be\label{eq.degroot1}
x(k+1)=W^{k}x(0)\xrightarrow[k\to\infty]{}(p_{\infty}^{\top}x(0))\ones_n.
\ee

The element $p_{\infty i}$ can thus be treated as a measure of \emph{social power} of agent $i$, i.e. the weight of its initial opinion $x_i(0)$ in the final opinion of the group. The greater this weight is, the more influential
is the $i$th individual's opinion. A more detailed discussion of social power and social influence mechanism is provided in~\citep{French:1956,Friedkin:1986}. The social power may be considered as a \emph{centrality measure}, allowing to identify the most ``important'' (influential) nodes of a social network. This centrality measure is similar to the \emph{eigenvector centrality}~\citep{Bonacich:01}, which is defined as the left eigenvector of the conventional binary adjacency matrix of a graph
instead of the ``normalized'' stochastic adjacency matrix. Usually centrality measures are introduced as functions of the graph topology~\citep{Newman:2003} while their relations to dynamical processes over graphs are not well studied. French's model of social power introduces a \emph{dynamic} mechanism of centrality measure and a decentralized algorithm~\eqref{eq.french1-dual} to compute it.
\begin{exam}
Consider the French model with $n=3$ agents~\eqref{eq.ex-1}, corresponding to the graph in Fig.~\ref{fig.french}. One can expect that the ``central'' node $2$ corresponds to the most influential agent in the group. This is confirmed by a straightforward computation: solving the system of equations $p_{\infty}^{\top}=p_{\infty}^{\top}W$ and $p_{\infty}^{\top}\ones=1$, one
obtains the vector of social powers $p_{\infty}^{\top}=(\tfrac{2}{7},\tfrac{3}{7},\tfrac{2}{7})$.
\end{exam}

\subsection{Stubborn agents in the French-DeGroot model}\label{subsec.stub}

  Although consensus is a typical behavior of the model~\eqref{eq.french}, there are situations when the opinions do not reach consensus but split into several clusters. One of the reasons for that is the presence of \emph{stubborn} agents (called also \emph{radicals}~\citep{HegselmannKrause:2015} or \emph{zealots}~\citep{Masuda:2015}).
\begin{defn} (Stubbornness)
An agent is said to be \emph{stubborn} if its opinion remains unchanged independent of the others' opinions.
\end{defn}

\begin{figure}
\center
\includegraphics[width=0.6\columnwidth]{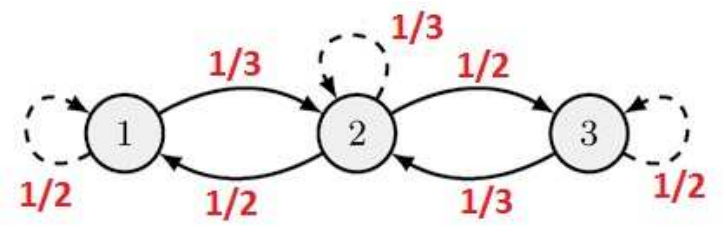}
\caption{The graph of the French-DeGroot model with two stubborn agents (source nodes) 1 and 3.}\label{fig.french-stub}
\end{figure}

If the opinions obey the model~\eqref{eq.french} then agent $i$ is stubborn $x_i(k)\equiv x_i(0)$ if and only if $w_{ii}=1$. Such an agent corresponds to a \emph{source} node in a graph $\g[W]$, i.e. a node having no incoming arcs but for the self-loop (Fig.~\ref{fig.french-stub}).
 Theorem~\ref{thm.degroot} implies that if $\g[W]$ has the only source, being also a \emph{root} (Fig.~\ref{fig.strong}a), then the opinions reach a consensus (the source node is the only closed strong component of the graph). If more than one stubborn agent exist (i.e. $\g[W]$ has several sources), then consensus among them is, obviously, impossible. Theorem~\ref{thm.degroot} implies, however, that typically the opinions in such a group converge.
\begin{cor}\label{cor.stubborn}
Let the group have $s\ge 1$ stubborn agents, influencing all other individuals (i.e. the set of source nodes
is connected by walks to all other nodes of $\g[W]$). Then the model~\eqref{eq.french} is convergent.
\end{cor}
Indeed, source nodes are the only closed strong components of $\g[W]$, which are obviously aperiodic.

In Section~\ref{sec.fj} it will be shown that under the assumptions of Corollary~\ref{cor.stubborn} the final opinion $x(\infty)$ is fully determined by the stubborn agents' opinions\footnote{This fact can also be derived from the Markov chain theory. In the dual Markov chain~\eqref{eq.french1-dual}, stubborn agents correspond to \emph{absorbing states}. The condition from Corollary~\ref{cor.stubborn} implies that all other states of the chain are non-recurrent, i.e. the Markov chain is \emph{absorbing}~\cite{KemenySnellMarkovBook} and thus arrives with probability $1$ at one of the absorbing states. Thus the columns of the limit matrix $W^{\infty}$, corresponding to non-stubborn agents, are zero.}.

\begin{exam}
Consider the French-DeGroot model, corresponding to the weighted graph in Fig.~\ref{fig.french-stub}
\[
\begin{bmatrix}
x_1(k+1)\\x_2(k+1)\\x_3(k+1)
\end{bmatrix}=\begin{pmatrix}
1 & 0 & 0\\
\frac{1}{3} & \frac{1}{3} & \frac{1}{3}\\
0 & 0 & 1
\end{pmatrix}\begin{bmatrix}
x_1(k)\\x_2(k)\\x_3(k)
\end{bmatrix}.
\]
It can be shown that the steady opinion vector of this model is $x(\infty)=(x_1(0),x_1(0)/2+x_3(0)/2, x_3(0))^{\top}$.
\end{exam}

%% file: 4abelson.tex
\section{Abelson's Models and Diversity Puzzle}\label{sec.abelson}

In his influential work~\citep{Abelson:1964} Abelson proposed a continuous-time counterpart of the French-DeGroot model~\eqref{eq.french}. Besides this model and its nonlinear extensions, he formulated a key problem in opinion formation modeling, referred to as the \emph{community cleavage} problem~\citep{Friedkin:2015} or Abelson's \emph{diversity puzzle}~\citep{Nakamura:16}.

\subsection{Abelson's models of opinion dynamics}\label{subsec.lin-abelson}

To introduce Abelson's model, we first consider an alternative interpretation of
the French-DeGroot model~\eqref{eq.french}. Recalling that $1-w_{ii}=\sum_{j\ne i}w_{ij}$, one has
\be\label{eq.french2}
\underbrace{x_i(k+1)-x_i(k)}_{\Delta x_i(k)}=\sum_{j\ne i}\underbrace{w_{ij}[x_j(k)-x_i(k)]}_{\Delta_{(j)} x_i(k)}\quad\forall i.
\ee
The experiments with dyadic interactions ($n=2$) show that ``the attitude positions of two discussants ... move toward each other''~\citep{Abelson:1967}. The equation~\eqref{eq.french2} stipulates that this argument holds for simultaneous interactions of multiple agents: adjusting its opinion $x_i(k)$ by $\Delta_{(j)} x_i(k)$, agent $i$ shifts it towards $x_j(k)$ as
\[
x_i'=x_i+\Delta_{(j)} x_i\Longrightarrow |x_j-x_i'|=(1-w_{ij})|x_j-x_i|.
\]
The increment in the $i$th agent's opinion $\Delta x_i(k)$ is the ``resultant'' of these simultaneous adjustments.

Supposing that the time elapsed between two steps of the opinion iteration is very small, the model~\eqref{eq.french2} can be replaced by
the continuous-time dynamics
\be\label{eq.abelson0}
\dot x_i(t)=\sum_{j\ne i}a_{ij}(x_j(t)-x_i(t)),\,i=1,\ldots,n.
\ee
Here $A=(a_{ij})$ is a non-negative (but not necessarily stochastic) matrix of infinitesimal influence weights (or ``contact rates''~\cite{Abelson:1964,Abelson:1967}).
The infinitesimal shift of the $i$th opinion $dx_i(t)=\dot x_i(t)dt$ is the superposition of the infinitesimal's shifts $a_{ij}(x_j(t)-x_i(t))dt$ of agent $i$'s towards the influencers.
A more general nonlinear mechanism of opinion evolution~\cite{Abelson:1964,Abelson:1967,HunterDanesCohenBook_1984} is
\be\label{eq.abelson1}
\dot x_i(t)=\sum_{j\ne i}a_{ij}g(x_i,x_j)(x_j(t)-x_i(t))\,\quad\forall i.
\ee
Here $g:\r\times\r\to (0;1]$ is a coupling function, describing the complex mechanism of opinion assimilation\footnote{The reasons to consider nonlinear couplings between the individuals opinions (attitudes) and possible types of such couplings are discussed in~\cite{HunterDanesCohenBook_1984}. Many dynamic models, introduced in~\cite{HunterDanesCohenBook_1984}, are still waiting for a rigorous mathematical analysis.}.

In this section, we mainly deal with the \emph{linear Abelson model}~\eqref{eq.abelson0}, whose equivalent matrix form is
\be\label{eq.abelson}
\dot x(t)=-L[A]x(t),
\ee
where $L[A]$ is the Laplacian matrix~\eqref{eq.laplacian}.
Recently the dynamics~\eqref{eq.abelson} has been rediscovered in multi-agent control theory~\citep{Murray:04,RenBeardBook,MesbahiEgerBook} as a continuous-time consensus algorithm. We discuss the convergence properties of this model in the next subsection.

\subsection{Convergence and consensus conditions}


Note that Corollary~\ref{cor.m-matrix}, applied to the M-matrix $L[A]$ and $\la_0=0$, implies that all Jordan blocks, corresponding to
the eigenvalue $\la_0=0$, are trivial and for any other eigenvalue $\la$ of the Laplacian $L[A]$ one has $\re\la>0$.
Thus, the model~\eqref{eq.abelson} is Lyapunov stable (yet not asymptotically stable) and, unlike the French-DeGroot model, is \emph{always} convergent.
\begin{cor}\label{cor.abelson}
For any nonnegative matrix $A$ the limit $P^{\infty}=\lim\limits_{t\to\infty}e^{-L[A]t}$ exists, and thus the vector of opinions in~\eqref{eq.abelson} converges $x(t)\xrightarrow[t\to\infty]{}x^{\infty}=P^{\infty}x(0)$.
\end{cor}
The matrix $P^{\infty}$ is a projection operator onto the Laplacian's null space $\ker L[A]=\{v: L[A]v=0\}$ and is closely related to the graph's structure~\cite{ChebotarevAgaev:2002,AgaevChe:2014}.

Similar to the discrete-time model~\eqref{eq.french}, the system~\eqref{eq.abelson} reaches a \emph{consensus} if the final opinions coincide $x^{\infty}_1=\ldots=x^{\infty}_n$ for any initial condition $x(0)$. Obviously, consensus means that $\ker L[A]$ is spanned by
the vector $\ones_n$, i.e. $P^{\infty}=\ones_np^{\top}_{\infty}$, where $p\in\r^n$ is some vector.
By noticing that $x=\ones_n$ is an equilibrium point, one has $P\ones_n=\ones_n$ and thus $p^{\top}_{\infty}\ones_n=1$.
Since $P$ commutes with $L[A]$, it can be easily shown that $p^{\top}_{\infty}L[A]=0$. Recalling that $L[A]$ has a \emph{nonnegative} left eigenvector $p$ such that $p^{\top}L[A]=0$ due to Corollary~\ref{cor.m-matrix} and $\dim\ker L[A]=1$, one has $p_{\infty}=cp$, where $c>0$. Combining this with Lemma~\ref{lem.laplacian}, one obtains the following consensus criterion.
\begin{thm}\label{thm.conse}
The linear Abelson model~\eqref{eq.abelson} reaches consensus if and only if $\g[A]$ is quasi-strongly connected (i.e. has a directed spanning tree). In this case, the opinions converge to the limit
\[
\lim_{t\to\infty}x_1(t)=\ldots=\lim_{t\to\infty}x_n(t)=p_{\infty}^{\top}x(0),
\]
where $p_{\infty}\in\r^n$ is the nonnegative vector, uniquely defined by the equations $p_{\infty}^{\top}L[A]=0$ and $p^{\top}_{\infty}\ones_n=1$.
\end{thm}
Similar to the French-DeGroot model, the vector $p_{\infty}$ may be treated as a vector of the agents' \emph{social powers}, or a centrality measure on the nodes of $\g[A]$.

It is remarkable that a crucial part of Theorem~\ref{thm.conse} was proved by Abelson~\cite{Abelson:1964}, who called quasi-strongly connected graphs ``compact''. Abelson proved that the null space $\ker L[A]$ consists of the vectors $c\ones_n$ if and only if the graph is ``compact'', i.e. statements 2 and 3 in Lemma~\ref{lem.laplacian} are equivalent. He concluded that ``compactness'' is necessary and sufficient for consensus; the proof, however, was given only for \emph{diagonalizable} Laplacian matrices. In general, the sufficiency part requires to prove that the zero eigenvalue of $L[A]$ is \emph{algebraically} simple (statement 1 in Lemma~\ref{lem.laplacian}).
The full proof of Theorem~\ref{thm.conse} was given only in~\citep{Ren:05}; the case of \emph{strong} graph was earlier considered in~\citep{Murray:04}.

As already discussed, the model~\eqref{eq.abelson} arises as a ``limit'' of the French-DeGroot model as the time between consecutive opinion updates becomes negligibly small.
The inverse operation of \emph{discretization} transforms~\eqref{eq.abelson} into the French-DeGroot model.
\begin{lem}~\citep{Ren:05,RenBeardBook}\label{lem.abelson-discrete}
For any nonnegative matrix $A$ and $\tau>0$ the matrix $W_{\tau}=e^{-\tau L[A]}$ is stochastic, and thus $P^{\infty}=\lim\limits_{\tau\to\infty}W_{\tau}$ from Corollary~\ref{cor.abelson} is stochastic. The matrices $W_{\tau}$ have positive diagonal entries.
\end{lem}

Lemma~\ref{lem.abelson-discrete} implies that the vectors $\tilde x(k)=x(\tau k)$ satisfy a special French-DeGroot model with $W=W_{\tau}$ and allows to derive Theorem~\ref{thm.conse} from Corollary~\ref{cor.harary}; this lemma can also be used for analysis of time-varying extensions of Abelson's model~\citep{RenBeardBook}.

Many results, regarding consensus algorithms over time-varying graphs, have been obtained in~\citep{Murray:04,RenBeardBook,MesbahiEgerBook,Moro:04,TsiTsi:13,MatvPro:2013,MartinGirard:2013,MartinHendrickx:2016,ProMatvCao:2016} and extended to general dynamic agents~\citep{RenBeardBook,RenCaoBook,LewisBook,ProCao16-EEEE,LiDuanChen:10}. More advanced results on \emph{nonlinear} consensus algorithms~\citep{LinFrancis:07,MatvPro:2013,Muenz:11} allow to examine the nonlinear Abelson model~\eqref{eq.abelson1} under different assumptions on the coupling function $g(\cdot)$. The statement of Abelson~\citep{Abelson:1964,Abelson:1967} that the model~\eqref{eq.abelson1} reaches consensus for any function $g(a,b)\in (0;1]$ when $\g[A]$ is ``compact'' (quasi-strongly connected) is, obviously, incorrect unless additional assumptions are adopted\footnote{Moreover, if the mapping $g(\cdot)$ is discontinuous, the system~\eqref{eq.abelson1} may have no solution in the classical sense.};
however, it holds for continuous function $g(a,b)$, as implied by the results of~\citep{LinFrancis:07,Muenz:11}.

\subsection{The community cleavage problem}\label{subsec.nonlin-abelson}

Admitting that in general the outcome of consensus is ``too strong to be realistic''~\citep{Abelson:1967}, Abelson formulated a fundamental problem, called
the \emph{community cleavage problem}~\citep{Friedkin:2015} or Abelson's \emph{diversity puzzle}~\citep{Nakamura:16}. The informal formulation, stated in~\citep{Abelson:1964}, was: ``Since universal ultimate agreement is an ubiquitous outcome of a very broad class of mathematical models, we are naturally led to inquire what on earth one must assume in order to generate the bimodal outcome of community cleavage studies.'' In other words, the reasons for \emph{social cleavage}, that is, persistent disagreement among the agents' opinions (e.g. clustering~\cite{XiaCao:11})
are to be identified. This requires to find mathematical models of opinion formation that are able to capture the complex behavior of real social groups, yet simple enough to be rigorously examined.

As discussed in Subsect.~\ref{subsec.stub}, one of the reasons for opinion clustering is the presence of \emph{stubborn} agents, whose opinions are invariant. In the models~\eqref{eq.abelson0} and \eqref{eq.abelson1}, agent $i$ is stubborn if and only if $a_{ij}=0\,\forall j$, corresponding thus to a source node of the graph $\g[A]$.
In the next sections we consider more general models with ``partially'' stubborn, or \emph{prejudiced}, agents. 

%% file: 5stubborn.tex
\section{Cleavage and Prejudices: Taylor's model}\label{sec.taylor}

In this section, we consider an extension of the linear Abelson model~\eqref{eq.abelson0}, proposed in~\citep{Taylor:1968}.
Whereas the French-DeGroot and Abelson models have triggered extensive studies on multi-agent consensus, Taylor's model in fact
has anticipated the recent studies on \emph{containment control} algorithms~\cite{JiEgerstedtBuffa:08,RenCaoBook,LiuXieWang:12,Cao:15-OpinionContainment}.

The model from~\citep{Taylor:1968}, as usual, involves $n$ agents with opinions $x_1,\ldots,x_n\in\r$ and $m\ge 1$ \emph{communication sources} (such as e.g. mass media), providing static opinions $s_1,\ldots,s_m\in\r$. The agents' opinions are influenced
by these sources, obeying the model
\be\label{eq.taylor0}
\dot x_i(t)=\sum_{j=1}^na_{ij}(x_j(t)-x_i(t))+\sum_{k=1}^mb_{ik}(s_k-x_i(t)).
\ee
Besides the nonnegative matrix of influence weights $A=(a_{ij})$, the Taylor model~\eqref{eq.taylor0} involves the
non-square $n\times m$ nonnegative matrix $B=(b_{ik})$ of ``persuasibility constants''~\citep{Taylor:1968}, which describe the influence of the communication sources on the agents. Some agents can be free of the external influence $b_{i1}=\ldots=b_{im}=0$, whereas the agents with $\sum_{k=1}^mb_{ik}>0$ are influenced by one or several sources.
Taylor has shown that the presence of external influence typically causes the cleavage of opinions; moreover,
unlike the Abelson model, the system~\eqref{eq.taylor0} is usually \emph{asymptotically stable} and converges to the unique equilibrium, determined by $s_1,\ldots,s_k$.

Besides the linear model~\eqref{eq.taylor0}, Taylor~\cite{Taylor:1968} considered
nonlinear opinion dynamics, which extend~\eqref{eq.abelson1} and some other models from~\citep{Abelson:1964} and
are still waiting for a rigorous mathematical examination. These systems are however beyond the scope of this tutorial.

\subsection{Equivalent representations of the Taylor model}

Note that formally the model~\eqref{eq.taylor0} may be considered as the Abelson model with $n+k$ agents, where the ``virtual'' agents $n+1,\ldots,n+k$ are \emph{stubborn}:
$x_{n+i}=s_i$ for $i=1,\ldots,k$. Corollary~\ref{cor.abelson} implies that the model~\eqref{eq.taylor0} is \emph{convergent}: for any $x(0)$ and $s_1,\ldots,s_k$ there
exist the limit $x(\infty)=\lim\limits_{t\to\infty}x(t)$. The converse is also true: for the Abelson model with $k\ge 1$
stubborn agents, their static opinions may be considered as ``communication sources'' for the others.
However, some properties of the system (e.g. stability) are easier to formulate for Taylor's system~\eqref{eq.taylor0} than for the augmented Abelson's model.

Another transformation allows to reduce~\eqref{eq.taylor0} to a formally less general model, where each agent may have
its own ``communication source'' or \emph{prejudice}
\be\label{eq.taylor0+}
\dot x_i(t)=\sum_{j=1}^na_{ij}(x_j(t)-x_i(t))+\gamma_i(u_i-x_i(t))
\ee
where $\gamma_i\ge 0$. Obviously,~\eqref{eq.taylor0} reduces to~\eqref{eq.taylor0+}, choosing $\gamma_i\dfb\sum_{m=1}^kb_{im}\ge 0$ and $u_i\dfb \gamma_i^{-1}\sum_{m=1}^kb_{im}s_m$ (if $\gamma_i=0$,
we set $u_i=0$ without loss of generality).
\begin{defn}(Prejudiced agents)
Given a group of $n$ agents, governed by the model~\eqref{eq.taylor0+}, we call agent $i$ \emph{prejudiced}
if $\gamma_i>0$; the external inputs $u_i$ are referred to as the \emph{prejudices} of
corresponding agents\footnote{Note that the model~\eqref{eq.taylor0+} has been studied in~\cite{XiaCao:11} as a protocol for multi-agent clustering;
prejudiced agents in~\cite{XiaCao:11} are called \emph{informed}, whereas other agents are said to be \emph{naive}.}.
\end{defn}

The prejudice may be considered as some ``internal'' agent's opinion, formed by some external factors (as in the Taylor model~\eqref{eq.taylor0}) or the individual's personal experience. An agent that is not prejudiced obeys the usual Abelson mechanism~\eqref{eq.abelson0}.
A prejudiced agent may be totally closed to the interpersonal influence $a_{ij}=0\,\forall j$; in this case
its opinion converges to its prejudice $x_i(t)\xrightarrow[t\to\infty]{} u_i$ and $\gamma_i$ regulates the
convergence rate. In the special case where $u_i=x_i(0)$ such an agent is \emph{stubborn} since $x_i(t)\equiv u_i$.
The concept of a prejudiced agent is however much more general and allows the agent to
be influenced by both its prejudice and the others' opinions.

\subsection{Stability of the Taylor model}

In this subsection, we examine asymptotic stability of the Taylor model. Obviously, it suffices to examine only the system~\eqref{eq.taylor0+}, to which the original model~\eqref{eq.taylor0} reduces.
Introducing the matrix $\Gamma=\diag(\gamma_1,\ldots,\gamma_n)$, the model~\eqref{eq.taylor0+} is rewritten as
\be\label{eq.taylor}
\dot x(t)=-(L[A]+\Gamma)x(t)+\Gamma u.
\ee

To examine the stability properties of~\eqref{eq.taylor}, we split the agents into two classes.
Agent $i$ is said to be \emph{P-dependent} (prejudice-dependent) if is either prejudiced or influenced
by some prejudiced agent $j$ (that is, a walk from $j$ to $i$ exists in the graph $\g[A]$).
Otherwise, we call the agent \emph{P-independent}.
Renumbering the agents, we assume that agents $1,\ldots,r\ge 1$ are P-dependent and agents $r+1,\ldots,n$ are P-independent
(possibly, $r=n$). Denote the corresponding parts of the opinion vector by, respectively, $x^1(t)$ and $x^2(t)$.
Since P-dependent agents are not connected to P-independent ones, Eq.~\eqref{eq.taylor} is decomposed as follows
\begin{align}
\dot x^1(t)&=-(L^{11}+\Gamma^{11})x^1(t)-L^{12}x^2(t)+\Gamma^{11} u^1\label{eq.taylor-d1}\\
\dot x^2(t)&=-L^{22}x^2(t)\label{eq.taylor-d2}.
\end{align}
The matrix $L^{22}$ is Laplacian of size $(n-r)\times (n-r)$, i.e.~\eqref{eq.taylor-d2} is the Abelson model.
The matrix $L^{11}$ is, in general, not Laplacian; one has $L^{11}\ones_r\ge 0$.
\begin{thm}\label{thm.taylor}
Let the community have $r\ge 1$ P-dependent agents and $n-r\ge 0$ P-independent ones. Then the dynamics of P-dependent agents~\eqref{eq.taylor-d1} is asymptotically stable, i.e.
the matrix $-(L^{11}+\Gamma^{11})$ is Hurwitz. The vector of their opinions converges to
\be\label{eq.taylor-opinions}
x^1(\infty)=M\begin{bmatrix}
u^1\\
x^2(\infty)
\end{bmatrix},\; M\dfb (L^{11}+\Gamma^{11})^{-1}\begin{bmatrix}
\Gamma^{11}\quad L^{12}\end{bmatrix}.
\ee
The matrix $M$ is stochastic, and thus the final opinion of any agent is a convex combination of the prejudices and the final opinions of P-independent agents\footnote{Recall that computation
of $x^2(\infty)$ reduces to the analysis of the Abelson model~\eqref{eq.taylor-d2}, discussed in Section~\ref{sec.abelson}.}.
\end{thm}

Theorem~\ref{thm.taylor} easily follows from the properties of M-matrices. Using Lemma~\ref{lem.m-matrix}, $L^{11}+\Gamma^{11}$ is proved to be M-matrix. It
suffices to show that the corresponding eigenvalue $\la_0$ from Corollary~\ref{cor.m-matrix} is positive.
Suppose on the contrary that $\la_0=0$ and let $p$ stand for the nonnegative left eigenvector $p^{\top}(L^{11}+\Gamma^{11})=0$. Multiplying by $\ones_r$ and noticing
that $L^{11}\ones_r\ge 0$, one has $p^{\top}\Gamma^{11}\ones_r=0$, that is, $p^{\top}\Gamma^{11}=0$ and $p_i=0$ whenever $\gamma_i>0$, i.e. $p_i=0$ for all prejudiced agents $i$.
Since $p^{\top}L^{11}=0$, for any $j$ such that $p_j=0$ one has $\sum_{i\ne j}p_ia_{ij}=p_j\sum_{j\ne i}a_{ji}=0$, i.e. $p_i=0$ whenever $a_{ij}>0$. In other words, if node $j$ is connected to node $i$ and $p_j=0$ then
also $p_i=0$. This implies that $p=0$ which contradicts to the choice of $p$. This contradiction shows that $\la_0>0$ and hence the system~\eqref{eq.taylor-d1} is stable, entailing~\eqref{eq.taylor-opinions}.
Since $(-L^{12})$ is nonnegative, Lemma~\ref{lem.m-matrix-inverse} implies that the matrix $M$ is also nonnegative. Choosing $u=\ones_n$, it is obvious that~\eqref{eq.taylor0+} has an equilibrium $x=\ones_n$, which implies that $M\ones_n=\ones_{r}$, i.e. $M$ is stochastic, which ends the proof.

\begin{cor}\label{cor.taylor}
The system~\eqref{eq.taylor0+} is asymptotically stable, i.e. the matrix $-(L[A]+\Gamma)$ is Hurwitz if and only if all agents are P-dependent.
\end{cor}

In terms of the original model~\eqref{eq.taylor0}, Corollary~\ref{cor.taylor} can be reformulated as follows:
the system~\eqref{eq.taylor0} is asymptotically stable if and only if any agent is influenced by at least
one ``communication source''\footnote{This statement was formulated in~\citep{Taylor:1968} (Theorem~1).}. This influence can be direct (if $\gamma_i=\sum_{m=1}^kb_{im}>0$) or
indirect (through a chain of the other agents).

\subsection{The Taylor model and containment control}\label{subsec.containment}

A multidimensional extension of the Taylor model~\eqref{eq.taylor0} arises in the \emph{containment control} problems~\cite{JiEgerstedtBuffa:08,RenCaoBook,LiuXieWang:12,Cao:15-OpinionContainment}.
The agents stand for mobile robots or vehicles, and the ``opinion'' $x_i\in\r^d$ is the position of agent $i$.
The ``communication sources'' $s_1,\ldots,s_k\in\r^d$ are the positions
of $k$ static \emph{leaders}. The mobile agents' goal is to reach the convex hull
\[
\mathcal S\dfb\left\{\sum_{m=1}^k\alpha_ms_m:\alpha_m\ge 0,\sum_{m=1}^k\alpha_m=1\right\}\subset\r^d
\]
spanned by the leaders $s_1,\ldots,s_k$.
Agent $i$ is directly influenced by the leader $m$ (i.e. $b_{im}>0$)
if it can measure its position (in general, none of the agents
can observe the whole set $\mathcal S$). Similar to the scalar case, such agents can be called ``prejudiced''.
The other agents can be either ``P-dependent'' (indirectly influenced by one or several leaders)
or ``P-independent''.

\begin{thm}\label{thm.contain}
The three conditions are equivalent
\begin{enumerate}
\item the system~\eqref{eq.taylor0} is Hurwitz stable;
\item any agent is influenced directly or indirectly by one of the leaders (P-dependent);
\item the mobile agents reach the target convex hull
\be\label{eq.contain}
x_i(\infty)\in\mathcal S\quad\forall i=1,\ldots,n
\ee
for any positions of the leaders $s_1,\ldots,s_k\in\r^d$ and the initial conditions $x_1(0),\ldots,x_n(0)\in\r^d$.
\end{enumerate}
\end{thm}

The equivalence $1\Longleftrightarrow 2$ is established by Theorem~\ref{thm.taylor}. Obviously, $3\Longrightarrow 1$: choosing $s_1=\ldots=s_k=0$, one has $\mathcal S=\{0\}$ and~\eqref{eq.contain}
is the asymptotic stability condition.
It remains to prove the implication $1\Longrightarrow 3$. In the scalar case $d=1$ it is immediate from Theorem~\ref{thm.taylor}. In general, let $v\in\r^d$ and $\tilde x_i(t)=v^{\top}x_i(t)$, $\tilde s_m=v^{\top}s_m$.
Then, obviously $\tilde x_i,\tilde s_m$ obey the scalar model~\eqref{eq.taylor0} and thus
$v^{\top}x_i(\infty)=\tilde x_i(\infty)\ge \min_m\tilde s_m=\min_{m}v^{\top}s_m=\min_{s\in \mathcal S}v^{\top}s$. Since $v$ is arbitrary, one has $x_i(\infty)\in\mathcal S$ for any $i$.

In the recent literature on containment control~\cite{JiEgerstedtBuffa:08,RenCaoBook,LiuXieWang:12,CaoEgerstedt:12,ShiHongJohansson:2012} Theorem~\ref{thm.contain} has been extended in various
directions: the leaders may be \emph{dynamic} (and thus their convex hull is time-varying $\mathcal S=\mathcal S(t)$), the interaction graph may also be time-varying and the agents
may have non-trivial dynamics. Furthermore, the polyhedron $\mathcal S$ can be replaced by an arbitrary closed convex set; the relevant problem is sometimes referred to as the \emph{target aggregation}~\cite{ShiHong:09}
and is closely related to distributed optimization~\cite{ShiJohanssonHong:13}.

\section{Friedkin-Johnsen Model}\label{sec.fj}

It is a remarkable fact that no discrete-time counterpart of the Taylor model had been suggested till 1990s, when Friedkin and Johnsen~\cite{FriedkinJohnsen:1990,FriedkinBook,FriedkinJohnsen:1999}
introduced a discrete-time modification of the dynamics~\eqref{eq.taylor0+}. Unlike many models of opinion formation, proposed in the literature, this model has been \emph{experimentally validated}
for small and medium-size groups, as reported in~\cite{FriedkinJohnsen:1999,Friedkin:2012,FriedkinJohnsenBook,Friedkin:2015,FriedkinJiaBullo:2016}.

Similar to DeGroot's dynamics~\eqref{eq.french}, the Friedkin-Johnsen model employs a stochastic matrix of social influences $W$, corresponding to the influence graph $\g[W]$. Besides this matrix, a
\emph{diagonal} matrix
$\Lambda=\diag(\la_{1},\ldots,\la_n)$ is introduced, where $\la_i\in [0,1]$ stands for the \emph{susceptibility} of agent $i$ to the process of social influence. The vector of the agents' opinions evolves in accordance with
\be\label{eq.fj}
x(k+1)=\La Wx(k)+(I-\La)u.
\ee
Here $u$ is a constant vector of the agents' \emph{prejudices}. The susceptibilities' complements $1-\la_i$ play the same role as the
coefficients $\gamma_i$ in~\eqref{eq.taylor0+}; in the case $\La=0$ the model~\eqref{eq.fj} turns into the French-DeGroot model~\eqref{eq.french}. If $1-\la_i=0$ then agent $i$ is
independent of the prejudice vector $u$ and applies the usual French-DeGroot ``opinion pooling'' rule. When $\la_i<1$, agent $i$ is ``anchored'' to its prejudice $u_i$ and factors
its into any opinion iteration. If $\la_i=0$ then the $i$th agent's opinion stabilizes at the first step $x_i(k)\equiv u_i\,\forall k\ge 1$; such an agent is \emph{stubborn}
$x_i(k)\equiv x_i(0)$ when $u_i=x_i(0)$.

In the Friedkin-Johnsen theory~\cite{FriedkinJohnsen:1999,FriedkinJohnsenBook,Friedkin:2015} it is supposed traditionally that $u=x(0)$, i.e. the prejudices of the agents are their initial opinions.
This is explained by the assumption~\cite{FriedkinJohnsen:1999} that the individuals prejudices have been formed by some exogenous conditions, which influenced the group in the past;
in this sense the vectors of prejudices $u$ and initial opinions $x(0)$ store the information about the group's history. The assumption $u=x(0)$ in turn motivates to adopt
the ``coupling condition'' $1-\la_i=w_{ii}$, stating that the levels of agents' ``anchorage'' to the initial opinions are determined by their self-confidence weights. At the same time,
similar to Taylor's model, the prejudice may be independent of the initial opinion $x(0)$ and caused by media or some other ``communication sources''. For this reason, we do not adopt these coupling conditions in this tutorial, allowing the prejudices and initial opinions to be independent; the same holds for the matrices $\La$ and $W$.

\subsection{Convergence and stability conditions}

Similar to the Taylor model, a generic Friedkin-Johnsen model is asymptotically stable, i.e. the substochastic matrix $\La W$ is Schur stable $\rho(\La W)<1$. This holds e.g.
when $\La<I$ or $\La\ne I$ and $\La W$ is irreducible (since an irreducible substochastic matrix is either stochastic or Schur stable~\cite{Meyer2000Book}). In this subsection, we give a necessary
and sufficient stability condition, similar to Theorem~\ref{thm.taylor} and established in~\cite{Parsegov2017TAC}. Henceforth we assume that $\La\ne I_n$ since otherwise~\eqref{eq.fj} reduces
to the French-DeGroot model~\eqref{eq.french}.

Following Section~\ref{sec.taylor}, we call agent $i$ \emph{prejudiced} if $\la_{i}<1$, i.e. the prejudice $u_i$ influences its opinion at each step $k$. Agent $i$ is \emph{P-dependent} if it is prejudiced or influenced by some prejudiced agent $j$, that is, a walk from $j$ to $i$ in the graph $\g[W]$ exists. Otherwise, the agent is \emph{P-independent}. Renumbering the agents, one may assume that agents $1,\ldots,r$ are P-dependent (where $r\ge 1$), whereas agents $r+1,\ldots,n$ are P-independent (it is possible that $r=n$, i.e. all agents are P-dependent). We denote the corresponding parts of the opinion vector by, respectively, $x_1(t)\in\r^r$ and $x_2(t)\in\r^{n-r}$.
Since P-independent agents are, by definition, not prejudiced ($\la_i=1$), the system~\eqref{eq.fj} is decomposed as follows
\begin{align}
x_1(k+1)&=\La^{11}[W^{11}x_1(k)+W^{12}x_2(k)]+(I-\La^{11})u_1\label{eq.fj-d1}\\
x_2(k+1)&=W^{22}x_2(k)\label{eq.fj-d2}.
\end{align}
Notice that $W^{22}$ is a stochastic matrix, i.e. the P-independent agents obey the French-DeGroot model.

\begin{thm}\label{thm.fj}
Let the community have $r\ge 1$ P-dependent agents and $n-r\ge 0$ P-independent ones.
Then, the subsystem~\eqref{eq.fj-d1} is asymptotically stable, i.e. $\La^{11}W^{11}$
is Schur stable $\rho(\La^{11}W^{11})<1$. The model~\eqref{eq.fj} is convergent if and only if $r=n$ or~\eqref{eq.fj-d2}
is convergent, i.e. $W^{22}$ is regular. In this case
\be\label{eq.fj-opinions}
\begin{gathered}
x^1(\infty)=V\begin{bmatrix}
u^1\\
x^2(\infty)
\end{bmatrix},\\
V\dfb (I_r-\La^{11}W^{11})^{-1}\begin{bmatrix}
I_r-\La^{11} & \La^{11}W^{12}\end{bmatrix}.
\end{gathered}
\ee
The matrix $V$ is \emph{stochastic}\footnote{Sometimes $V$ is referred to as the \emph{control matrix}~\cite{Friedkin:2015}.}, i.e. the final opinion of any agent
is a convex combination of the prejudices and final opinions of P-independent agents\footnote{Recall that computation of $x^2(\infty)$ reduces to the analysis of the French-DeGroot model~\eqref{eq.fj-d2}, discussed in Section~\ref{sec.french}.}.
\end{thm}

Below we give the sketch of the proof of Theorem~\ref{thm.fj}, retracing the proof of Theorem~\ref{thm.taylor}.
An equivalent proof in~\cite{Parsegov2017TAC} relies on some properties on substochastic matrices\footnote{Note that~\cite{Parsegov2017TAC} uses
a different terminology: prejudiced agents are called ``stubborn'', stubborn agents in our sense are called ``totally stubborn'',
P-independent agents are called ``oblivious'', for P-dependent agents no special term is used.}.
Suppose on the contrary that $\rho(\La^{11}W^{11})=1$ and let $p\in\r^r$ stand
for the nonnegative left eigenvector, corresponding to this eigenvalue $p^{\top}\La^{11}W^{11}=p^{\top}$ and such that
$p^{\top}\ones_r=1$. Since $W^{11}$ is substochastic, one has $p^{\top}\La^{11}\ones_r\ge 1$ and thus $p_i=0$
when $\la_i<1$ (i.e. agent $i$ is prejudiced).
Recalling that $p$ is a left eigenvector, one has $p_j\la_j=\sum_{i=1}^rp_i\la_iw_{ij}$ for any $j=1,\ldots,r$ and thus
if $p_j=0$ and $w_{ij}>0$ (i.e. $j$ is connected to $i$) then $p_i\la_i=0$, which implies that $p_i=0$
(as we have already shown, $\la_i=0$ entails that $p_i=0$). Thus $p=0$, which is a contradiction and
thus $\La^{11}W^{11}$ is Schur stable. The second statement and the validity of~\eqref{eq.fj-opinions} if~\eqref{eq.fj-d2}
converges are now obvious. To prove that $V$ is stochastic, note first that $V\ones_n=\ones_r$.
Indeed, $(I_r-\La^{11})\ones_r+\La^{11}W^{12}\ones_{n-r}=(I_r-\La^{11})\ones_r+\La^{11}(\ones_r-W^{11}\ones_r)=
(I_r-\La^{11}W^{11})\ones_r$. On the other hand, $(I_r-\La^{11}W^{11})$ is an M-matrix, and thus $V$ is
nonnegative thanks to Lemma~\ref{lem.m-matrix-inverse}.

\begin{cor}\label{cor.fj-1}
The Friedkin-Johnsen model~\eqref{eq.fj} is asymptotically stable if and only if all agents are P-dependent.
Then, $V=(I-\La W)^{-1}(I-\La)$.
This holds, in particular, if $\La<I_n$ or $\La\ne I_n$ and $\g[W]$ is strongly connected ($W$ is irreducible).
\end{cor}

Corollary~\ref{cor.fj-1} may be transformed into a criterion of Schur stability for substochastic matrices since
the matrix $A$ is substochastic if and only if $A=\La W$ with diagonal $\La$ (where $0\le \La\le I$) and stochastic $W$.
The sufficiency part in Corollary~\ref{cor.fj-1} was proved in~\cite{FrascaTempo:2013}.

\begin{cor}\label{cor.fj-2}
The model~\eqref{eq.fj} converges if and only if $A=\La W$ is regular, i.e.
the limit $\lim\limits_{k\to\infty}A^k$ exists.
\end{cor}

To prove Corollary~\ref{cor.fj-2} it remains to notice that $A$ is regular if and only if its submatrix
$W^{22}$ from~\eqref{eq.fj-d2} is regular. This result is formulated in~\cite{FriedkinJohnsen:1999,Friedkin:2015} without rigorous proof. The property from Corollary~\ref{cor.fj-2}
is non-trivial since in general the system
\[
x(k+1)=Ax(k)+Bu
\]
with a regular matrix $A$ and some matrix $B$ is not convergent and may have unbounded solutions,
as demonstrated by the counterexample $A=B=I$.

\begin{figure}[h]\center
\begin{subfigure}[b]{0.75\linewidth}
\center
  \includegraphics[height=4cm, width=7cm]{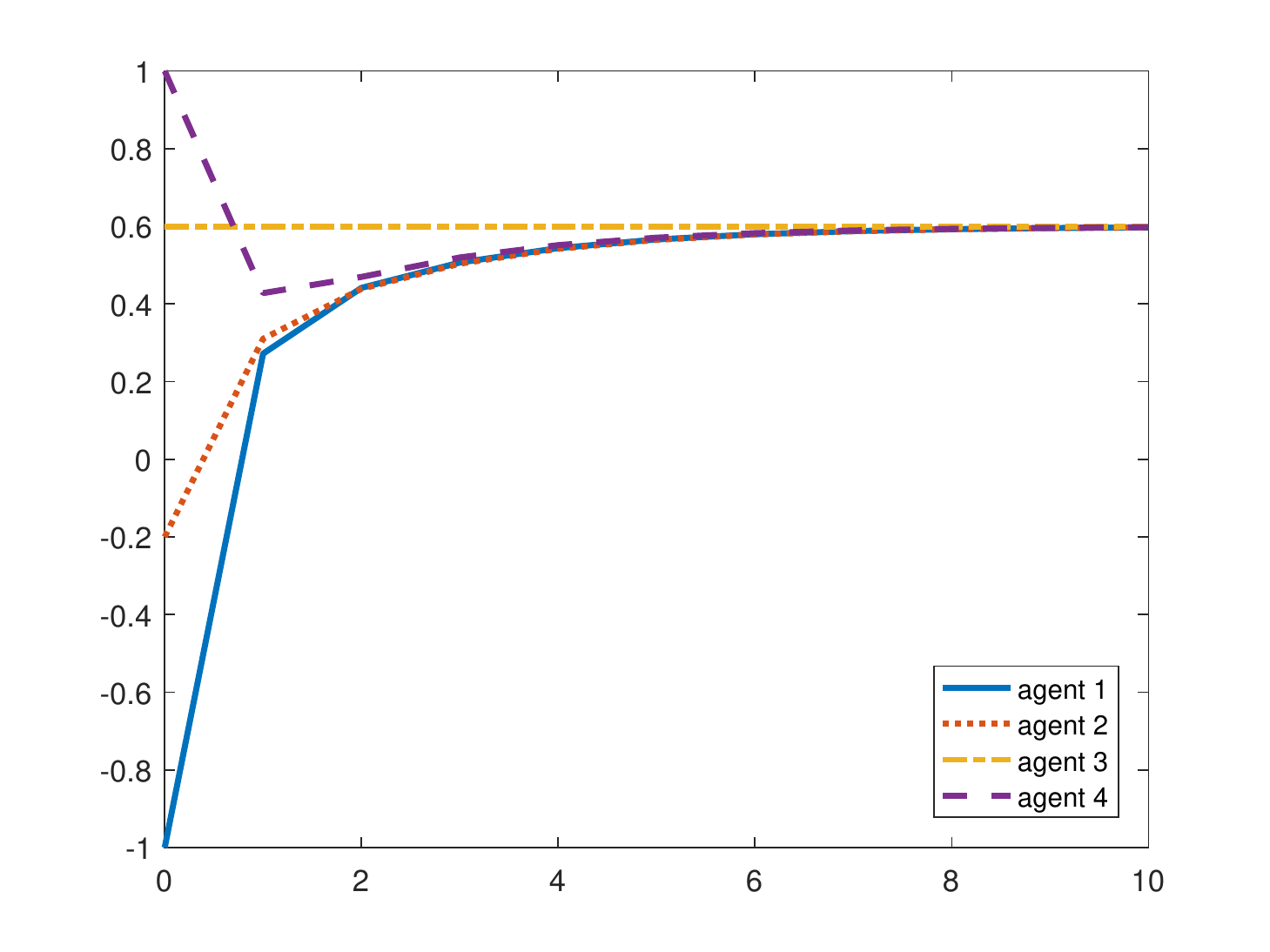}
\caption{$\La=I$}
 \end{subfigure}
 \begin{subfigure}[b]{0.75\linewidth}
 \center
  \includegraphics[height=4cm, width=7cm]{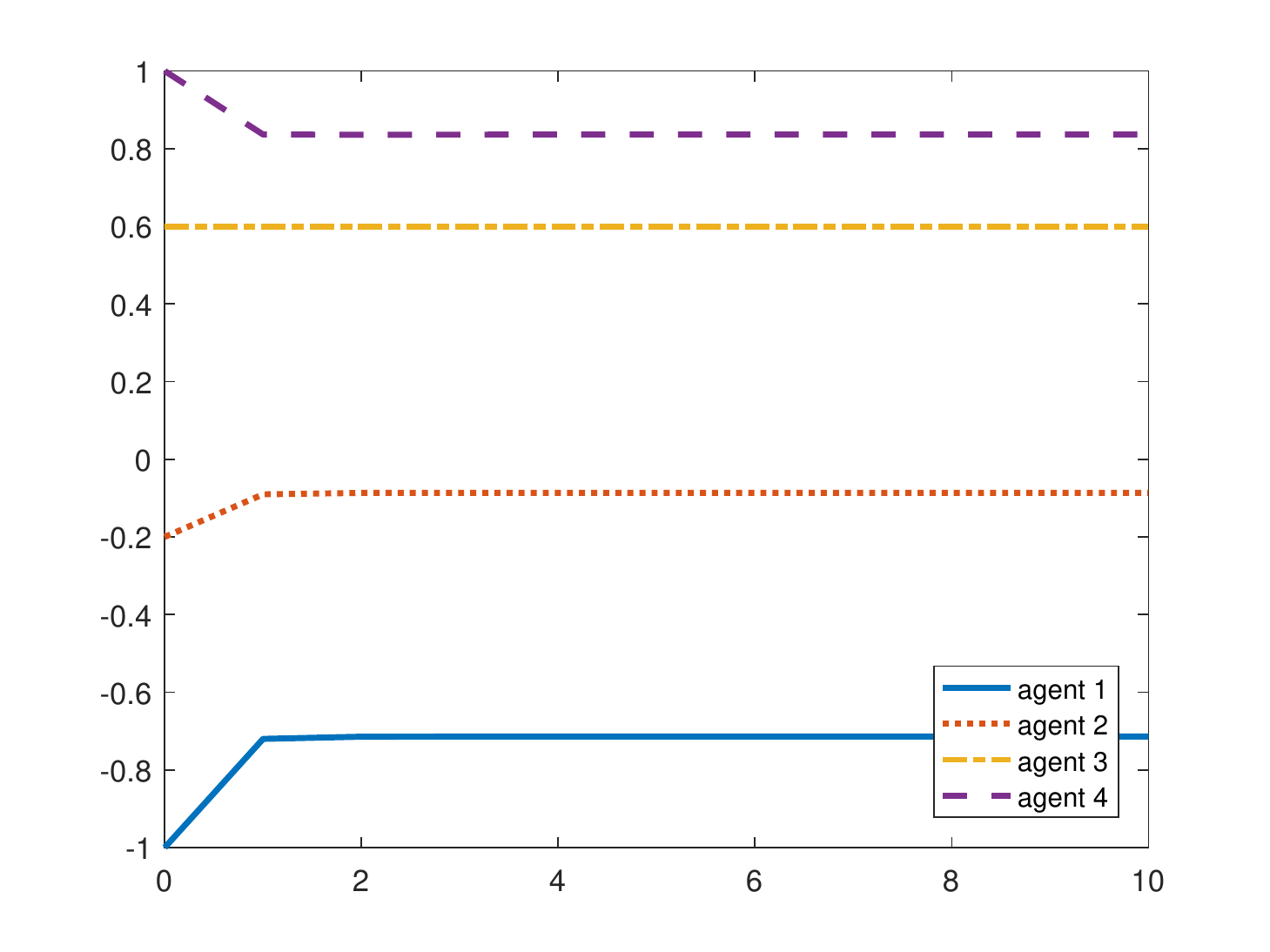}
     \caption{$\La=I-\diag(W)$}
 \end{subfigure}
  \begin{subfigure}[b]{0.75\linewidth}
  \center
  \includegraphics[height=4cm, width=7cm]{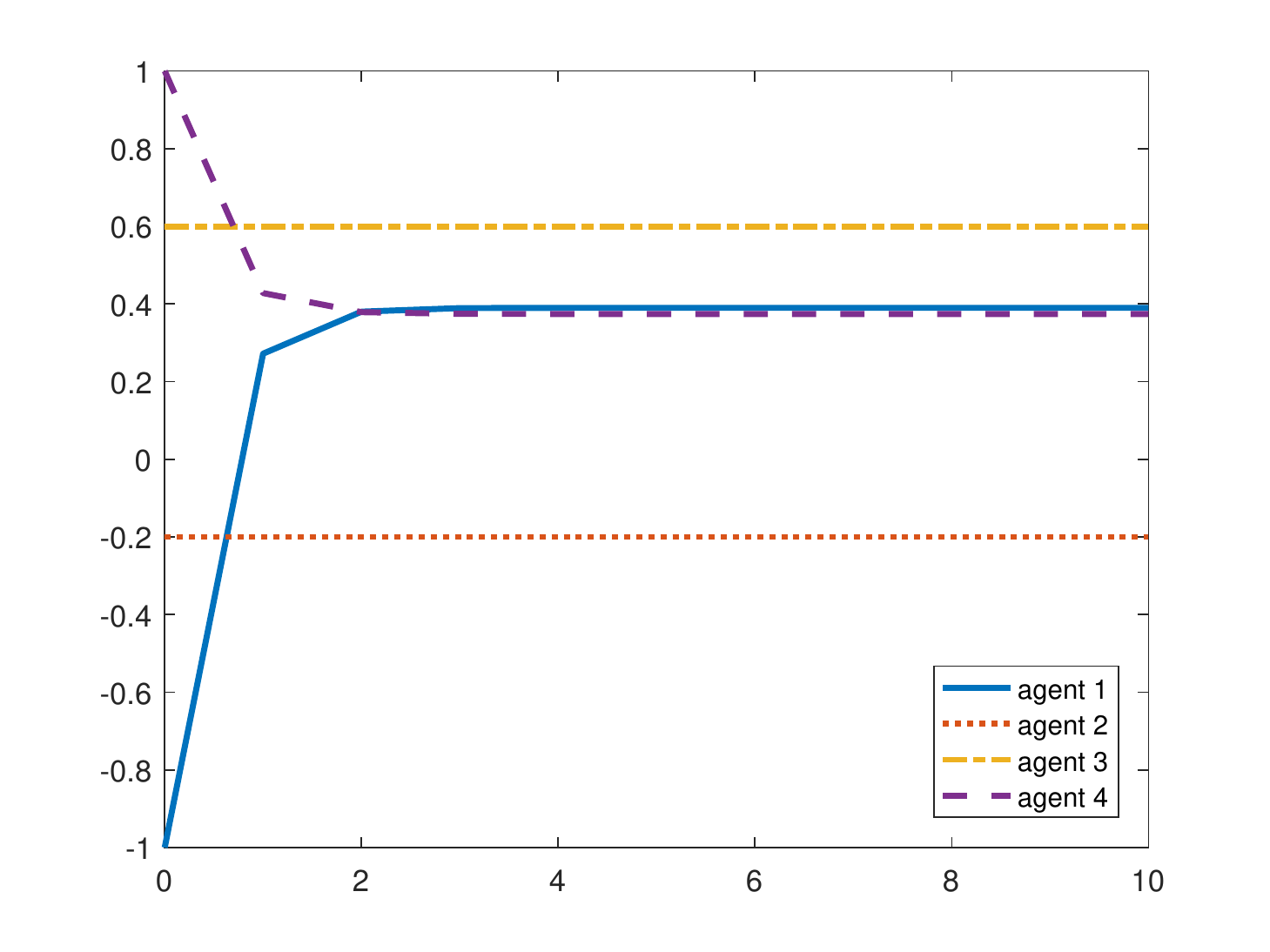}
      \caption{$\La=\diag(1,0,0,1)$}
 \end{subfigure}
 \caption{Opinion dynamics for $W$ from~\eqref{eq.W} and different $\La$.}\label{fig.fj}
 \end{figure}

\begin{exam}
This example illustrates the behavior of opinions in the Friedkin-Johnsen model~\eqref{eq.fj} with $n=4$ agents and the matrix of influence weights~\cite{FriedkinJohnsen:1999}
\be\label{eq.W}
  W = \begin{pmatrix}
  0.220 & 0.120 & 0.360 & 0.300 \\
   0.147 & 0.215 & 0.344 & 0.294 \\
   0 & 0 & 1 & 0 \\
   0.090 & 0.178 & 0.446 & 0.286
\end{pmatrix}.
\ee
We put $u=x(0)=[-1,-0.2,0.6,1]^{\top}$ and consider the evolution of opinions for three different matrices $\La$ (Fig.~\ref{fig.fj}): $\La=I$, $\La=I-\diag(W)$ and $\La=\diag(1,0,0,1)$. 
In all cases agent $3$ is stubborn. In the first case the model~\eqref{eq.fj} reduces to the French-DeGroot model, and the opinions reach consensus (Fig.~\ref{fig.fj}a). 
In the second case (Fig.~\ref{fig.fj}b) agents $1,2,4$ move their opinions towards the stubborn agent $3$'s opinion, however, the visible cleavage of their opinions is observed.
In the third case (Fig.~\ref{fig.fj}c) agents $2$ and $3$ are stubborn, and the remaining agents $1$ and $4$ converge to different yet very close opinions, 
lying between the opinions of the stubborn agents.
\end{exam}

\subsection{Friedkin's influence centrality and PageRank}

A natural question arises whether the concept of social power, introduced for the French-DeGroot model, can be extended to the model~\eqref{eq.fj}.
In this subsection, we discuss such an extension, introduced by Friedkin~\cite{Friedkin:1991,Friedkin:2015} and based on the equality~\eqref{eq.fj-opinions}.
We confine ourselves to the case of asymptotically stable model~\eqref{eq.fj} with the prejudice vector $u=x(0)$, hence $x(\infty)=Vx(0)$, where $V=(I-\La W)^{-1}(I-\La)$.

Recall that the definition of French's social power assumed that the agents converge to the same consensus opinion $x_1(\infty)=\ldots=x_n(\infty)$; the social power
of agent $i$ is defined as the weight of its initial opinion $x_i(0)$ in this final opinion of the group. The Friedkin-Johnsen model provides a generalization of social 
power of agent $i$ as the \emph{mean weight} of its initial opinion in determining group members' final opinions~\cite{Friedkin:2015}. Mathematically, these mean influence weights are elements
of the non-negative vector $c\dfb n^{-1}V^{\top}\ones_n$, which satisfies the following equality
\be\label{eq.x-bar}
\bar x=\frac{1}{n}\sum_{i=1}^nx_i(\infty)=\frac{1}{n}\ones_n^{\top}Vx(0)=c^{\top}x(0).
\ee
Following~\cite{Friedkin:1991,Friedkin:2015}, we call $c_i$ the \emph{influence centrality} of agent $i$. Since $V$ is stochastic,
$c^{\top}\ones_n=1$. Similar to French's social power, the influence centrality is generated by an opinion formation
mechanism\footnote{Notice that analogous
influence centrality can be introduced for Taylor's model~\eqref{eq.taylor0+} with the prejudice $u=x(0)$.
}.

Choosing a stochastic matrix $W$, adapted to a given graph $\g$ with $n$ nodes, and some diagonal matrix $\La$, Friedkin's construction
gives a very insightful and broad class of centrality measures. For a fixed $W$, let $\La=\alpha I_n$
with $\alpha\in (0,1)$ and consider the corresponding matrix $V_{\alpha}$ and vector $c_{\alpha}$. Obviously,
$c_0=n^{-1}\ones_n$, i.e. the social power is uniformly distributed between all agents.
As $\alpha\to 1$, the vector $c_{\alpha}$ converges to French's social power (when it exists).
\begin{lem}
Let $W$ be a fully regular matrix and $p_{\infty}$ stand for the vector of French's social power,
corresponding to the model~\eqref{eq.french}. Then, $p_{\infty}=\lim\limits_{\alpha\to 1-0} c_{\alpha}$.
\end{lem}

Indeed, as follows from~\cite[Eq. (12)]{Parsegov2017TAC}
$V_{\alpha}=(1-\alpha)(I-\alpha W)^{-1}\xrightarrow[\alpha\to 1-]{}W^{\infty}$, where
$W^{\infty}=\lim\limits_{k\to\infty}W^k=\ones_np^{\top}$.
Thus $c_{\alpha}=n^{-1}V^{\top}_{\alpha}\ones_n\xrightarrow[\alpha\to 1-]{} p_{\infty}$.

The class of Friedkin's centrality measures $f_{\alpha}$ contains the well-known \emph{PageRank}~\cite{BrinPage:1998,Boldi:09,IshiiTempo:2010,IshiiTempo:2014,FrascaIshiiTempo:2015,FrascaTempo:2015,YouTempoQui:2017}, 
proposed originally for ranking webpages in Web search engines and used in scientometrics for journal ranking~\cite{IshiiTempo:2014}. 
The relation between the influence centrality and PageRank, revealed in~\cite{FriedkinJohnsen:2014}, follows from the construction of PageRank via ``random
surfing''~\cite{Boldi:09}.

Consider a segment of World Wide Web (WWW) with $n$ webpages and a stochastic $n\times n$ \emph{hyperlink}
matrix $W=(w_{ij})$, where $w_{ij}>0$ if and only if a hyperlink leads from the $i$th webpage to the $j$th one\footnote{Usually,
$W$ is obtained from the adjacency matrix of some known web graph via normalization and removing
the ``dangling'' nodes without outgoing hyperlinks~\cite{Boldi:09,IshiiTempo:2014}.}. Reaching webpage $i$, the surfer randomly
follows one of the hyperlinks on it; the probability to choose the hyperlink leading to webpage $j$ is $w_{ij}$ (the webpage may refer to itself $w_{ii}>0$).
Such a process of random surfing is a Markov process; denoting the probability to open webpage $i$ on step $k$
by $p_i(k)$, the row vector $p(k)^{\top}=(p_1(k),\ldots,p_n(k))$ obeys the ``dual'' French-DeGroot
model~\eqref{eq.french1-dual}. As discussed in Section~\ref{sec.fj}, if the French social power
vector $p_{\infty}$ is well-defined, then the probability distribution $p(k)$ converges to $p_{\infty}$.
Since $p_{\infty}^{\top}=p^{\top}W$, the vector $p_{\infty}$ satisfies the natural principle of webpage ranking:
a webpage referred by highly ranked webpages should also get a high rank. However, the web
graphs are often disconnected, so $W$ may be not fully regular, i.e. the French's social power may not exist.
To avoid this problem, the Markov process of random surfing is modified, allowing
the \emph{teleportation}~\cite{Boldi:09} from each node to a randomly chosen webpage.
With probability $m\in (0,1)$, at each step the surfer ``gets tired'' and opens a random webpage, sampled from
the uniform distribution\footnote{Note however
that the procedure of journal ranking, which is beyond this tutorial, uses a non-uniform distribution~\cite{IshiiTempo:2014}.}.
The Markov chain~\eqref{eq.french1-dual} is replaced by
\be
p(k+1)^{\top}=(1-m)p(k)^{\top}W+\frac{m}{n}\ones_n^{\top},
\ee
which is dual to the Friedkin-Johnsen model~\eqref{eq.fj} with $\La=(1-m)I_n$. It can be easily shown that
$p(k)^{\top}\to c_{1-m}^{\top}=\frac{m}{n}\ones_n^{\top}(1-(1-m)W)^{-1}$. Being a special case
of Friedkin's influence centrality, this limit probability is referred to as \emph{PageRank}~\cite{Boldi:09,IshiiTempo:2014}
(the Google algorithm~\cite{BrinPage:1998} used the value $m=0.15$). Another extension of the PageRank centrality, based
on the general model~\eqref{eq.fj}, has been proposed in~\cite{ProTempoCao16-1}.

\subsection{Alternative interpretations and extensions}

Obviously, the French-DeGroot model~\eqref{eq.french} can be considered as a degenerate case of~\eqref{eq.fj}
with $\La=I_n$. The French-DeGroot model with stubborn agents, examined in Subsect.~\ref{subsec.stub},
may be transformed to the more insightful model~\eqref{eq.fj}, where $\la_i=0$ and $u_i=x_i(0)$ when agent $i$ is stubborn ($w_{ii}=1$)
and $\la_i=1$, $u_i=0$ otherwise. Under the assumption of Corollary~\ref{cor.stubborn}, the system~\eqref{eq.fj}
is asymptotically stable\footnote{It may seem paradoxical that the equivalent transformation of the neutrally stable French-DeGroot
model into~\eqref{eq.fj} yields in an asymptotically stable system. The explanation is that changing
the initial condition $x(0)$ in~\eqref{eq.fj}, the prejudice vector $u$ remains constant.
In the original system~\eqref{eq.french} the prejudice is a part of the state vector, destroying the asymptotic
stability.} and the final opinion vector $x(\infty)=Vu$ is determined by the stubborn agents' opinions.
On the other hand,~\eqref{eq.fj} may be considered as an ``augmented'' French-DeGroot model with ``virtual''
$n$ stubborn agents, anchored at $x_{n+i}\equiv u_i$ (here $i=1,\ldots,n$).

The model~\eqref{eq.fj} has an elegant game-theoretic interpretation~\cite{Bindel:2011,GhaderiSrikant:2014}.
Suppose that each agent is associated to a cost function $J_i(x)=\la_i\sum_{j=1}^nw_{ij}(x_j-x_i)^2+(1-\la_i)(x_i-u_i)^2$:
the first term penalizes the disagreement from the others' opinions, whereas the other term ``anchors''
the agent to its prejudice. The update rule~\eqref{eq.fj} implies that each agent $i$ updates its opinion
in a way to minimize $J_i(x)$, assuming that $x_j(k)$, $j\ne k$, are constant. If the system is convergent,
the vector $x(\infty)$ stands for the \emph{Nash equilibrium}~\cite{Bindel:2011} in the game, which however
does not optimize the overall cost functional $J(x)=\sum_i J_i(x)$. Some estimates for the ratio
$\frac{J(x(\infty))}{\min_x J(x)}$ (considered as the ``price of anarchy'') are given in~\cite{Bindel:2011}.
In some special cases the model~\eqref{eq.fj} can also be interpreted in terms of electric circuits~\cite{GhaderiSrikant:2014,FrascaIshiiTempo:2015}.

The model~\eqref{eq.fj} with scalar opinions can be extended to $d$-dimensional opinions; similarly to
the DeGroot model~\eqref{eq.degroot}, these opinions can be wrapped into a $n\times d$ matrix $X(k)$, whose $i$th row
$x^i(k)$ represents the opinion of agent $i$; the multidimensional prejudices are represented by the matrix $U$.
The process of opinion formation is thus governed by the model
\be\label{eq.fj-mult}
X(k+1)=\La WX(k)+(I-\La)U.
\ee
Similar to the multidimensional Taylor model in Subsect.~\ref{subsec.containment}, the model~\eqref{eq.fj-mult} may be
considered as a discrete-time containment control algorithm~\cite{MuYang:16}.
An important extension of the model~\eqref{eq.fj-mult}, proposed in~\cite{Parsegov2017TAC,FriedkinPro2016}
considers the case where an individual vector-valued opinion represents his/her positions on
several \emph{interdependent} topics (such opinions can stand e.g. for
\emph{belief systems}, obeying some logical constraints~\cite{Converse:1964,FriedkinPro2016}).
The mutual dependencies between the topics can be described by introducing additional ``internal'' couplings, described
by a stochastic $d\times d$ matrix $C$. The model~\eqref{eq.fj-mult} is replaced by
\be\label{eq.fj-mult1}
X(k+1)=\La WX(k)C^{\top}+(I-\La)U.
\ee
As shown in~\cite{Parsegov2017TAC}, the stability conditions for~\eqref{eq.fj-mult1} remain
the same as for the original model~\eqref{eq.fj}. In~\cite{FriedkinPro2016} (see also Supplement to this paper)
an extension of~\eqref{eq.fj-mult1} has been also examined, where agents' have heterogeneous sets of logical constraints,
corresponding to $n$ different coupling matrices $C_1,\ldots,C_n$.

Finally, gossip-based versions of the models~\eqref{eq.fj} and~\eqref{eq.fj-mult1} with asynchronous communication have been proposed
in~\cite{FrascaTempo:2013,FrascaTempo:2015,FrascaIshiiTempo:2015,Parsegov2017TAC}; these models will be considered in
Part II of this tutorial.